\documentclass[preprint,12pt]{elsarticle}

\usepackage{graphicx}
\usepackage{amssymb}

\usepackage{amsmath}
\usepackage{booktabs}
\usepackage{amssymb}
\usepackage{xcolor}
\usepackage[linesnumbered,ruled,vlined]{algorithm2e}
\usepackage{mathrsfs}

\journal{arxiv.org}

\usepackage{booktabs}
\begin{document}

\begin{frontmatter}


\title{Weakly Complete Semantics Based on\\ Undecidedness Blocking.}



\author{Pierpaolo Dondio, Luca Longo}

\address{School of Computer Science, Technological University Dublin pierpaolo.dondio@tudublin.ie, luca.longo@tudublin.ie}

\begin{abstract}
In this paper we introduce a novel family of semantics called \textit{weakly complete} semantics. Differently from Dung's \textit{complete} semantics, \textit{weakly complete} semantics employs a mechanism called \textit{undecidedness blocking} by which the label undecided of an attacking argument is not always propagated to an otherwise accepted attacked argument. The new semantics are conflict-free, non-admissible but employing a weaker notion of admissibility; they allow reinstatement and they retain the majority of properties of \textit{complete} semantics. 
We show how both \textit{weakly complete} and Dung's \textit{complete} semantics can be generated by applying different \textit{undecidedness blocking} strategies, making \textit{undecidedness blocking} a unifying mechanism underlying argumentation semantics.
The semantics are also an example of ambiguity blocking Dunganian semantics and the first semantics to tackle the problem of \textit{self-defeating} attacking arguments. 
In the last part of the paper we compare \textit{weakly complete} semantics with the recent work of Baumann et al. on \textit{weakly admissible} semantics.
Since the two families of semantics do not coincide, a principle-based analysis of the two approaches is provided. The analysis shows how our semantics satisfy a number of principles satisfied by Dung's \textit{complete} semantics but not by Baumann et al. semantics, including \textit{directionality}, \textit{abstention}, \textit{SCC-decomposability} and \textit{cardinality} of extensions, making them a more faithful non-admissible version of Dung’ semantics.
\end{abstract}


\end{frontmatter}


\section{Introduction}
\label{S:1}
Abstract argumentation is a framework for non-monotonic reasoning where conclusions are reached by evaluating arguments and their conflict relation. The formalism is centred on the notion of argumentation framework \cite{dung}, a directed graph where nodes represent arguments and links represent an attack relation defined over arguments. 
One of the main tasks of abstract argumentation is the computation of the acceptability status of the arguments composing the framework. This is performed by the application of an argumentation semantics which identifies the sets of arguments, called \textit{extensions}, which successfully survive the conflicts encoded in the attack relation. 
In the labelling approach \cite{caminada} adopted in this paper the effect of an argumentation semantics is to assign to each argument a label \texttt{in}, \texttt{out} or \texttt{undec}. This means that an argument can respectively be accepted, rejected or deemed undecided. The \texttt{undec} label represents a situation in which the semantics has no reasons to definitely accept or reject an argument.

This paper extends the studies of abstract argumentation by presenting a new family of abstract semantics, called \textit{weakly complete} semantics. 
While in Dung's \textit{complete} semantics the effect of an attack by an undecided argument $b$ to an otherwise accepted argument $a$ is always to deem $a$ undecided - therefore propagating the $undec$ label of $b$ to $a$ - in \textit{weakly complete} semantics the propagation of the \texttt{undec} label could be blocked by the postulates of the semantics, and $a$ could be accepted. We refer to this mechanism as \textit{undecidedness blocking}.
We perform a principle-based analysis of such new semantics and a comparison with the \textit{weakly admissible} semantics recently proposed by Baumann et al. \cite{bbu1}.

\begin{figure}[h]
\centering\includegraphics[width=0.5\linewidth]{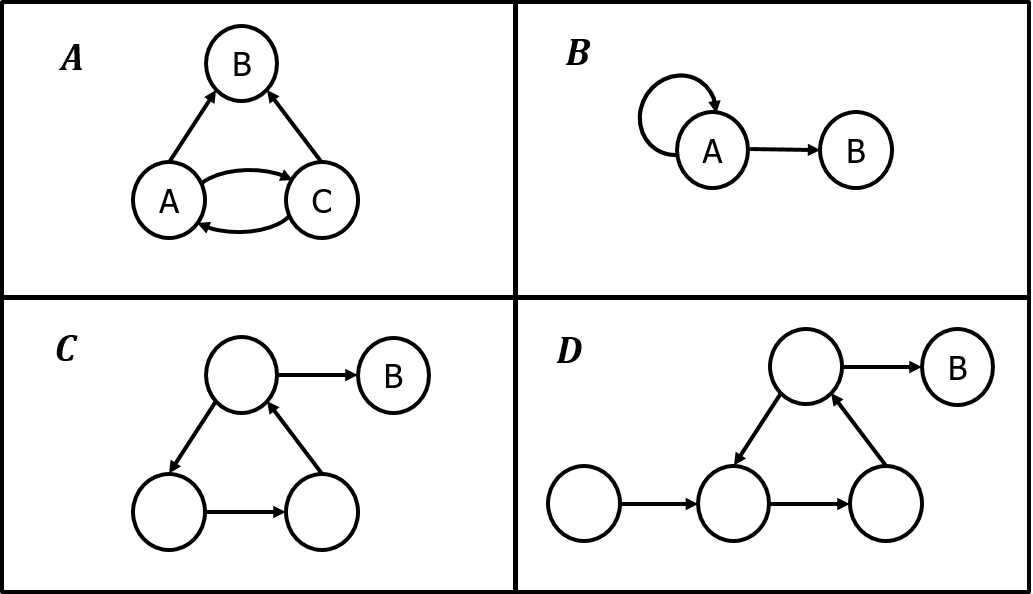}
\caption{Motivating examples. Graph A is an example of \textit{floating assignment} graph. Graph B shows the problem of self-defeating attacking arguments. Graph C is a variation of graph B where the attacking argument is part of odd-length cycle. In Graph D an attacker is added to the three-argument cycle of graph C.}
\end{figure}

The motivations for these new semantics are two: (1) to model the fundamental reasoning mechanism of \textit{ambiguity blocking} absent in current abstract semantics and (2) to propose semantics able to handle the 25-year old problem of self-defeating attacking arguments, referred in this paper as the problem of \textit{weak attacks}.

\vspace{2mm}
\noindent \textit{Motivation 1: Ambiguity Blocking and Abstract Argumentation}
\vspace{2mm}

\textit{Ambiguity blocking} semantics are employed by several non-monotonic formalisms, such as defeasible logic.
Informally, \textit{ambiguity blocking} is a mechanism by which statements for which there are contradictory evidence about their validity are not used to derive any conclusion and they do not affect the validity of other statements. An ambiguity blocking mechanism is certainly the most appropriate in a legal dispute (\cite{calegari2019defeasible}): if evidence versus an accused is not definitive or open to multiple interpretations, then evidence is void and the judge rules in favour of the accused. Referring to graph A in figure 1, using \textit{ambiguity blocking} there is a case for argument $b$ to be accepted, since $b$ is attacked by the two arguments $a$ and $c$ that are conflicting and therefore ambiguous, and therefore $a$ and $c$ is  rejected and $b$ accepted. The ambiguity of $a$ and $c$ is not propagated to $b$ and their attacks are \textit{de facto} neglected. Note how the same mechanism could be used to build a case for $b$ to be accepted in all the graphs of figure 1.

However, the translation of the \textit{ambiguity blocking} mechanism into abstract argumentation is not straightforward. The notion of ambiguity is formalized in defeasible logic, a rule-based system less abstract than Dung's framework. In defeasible logic knowledge is represented by rules over literals and inferences are obtained by chaining valid rules. An acyclic superiority relation is defined to model priorities among rules. In $DL$ a literal is \textit{ambiguous} if there is a valid chain of reasoning for $a$ and a valid one for $\lnot a$ and the superiority relation does not solve the conflict.
Both \textit{ambiguity blocking} and \textit{ambiguity propagating} semantics can be accommodated in $DL$. In the former ambiguous literals are rejected and their ambiguity is not transferred to other literals, while in the latter the ambiguity of the two literals is propagated to other related literals and rules.
Previous researches (\cite{van2012relating},\cite{governatori2004argumentation}) highlighted how Dung' semantics do not allow for an easy translation of the \textit{ambiguity blocking} mechanism. 
Indeed, the notion of \textit{ambiguity} in Dung’s framework is not even defined. Candidate to represent ambiguous arguments could be undecided arguments or \textit{credulously} accepted arguments but, as we discuss in a dedicated section of this paper, the concepts do not coincide.
\cite{governatori2004argumentation} showed how \textit{grounded} semantics can indeed instantiate \textit{ambiguity propagating} semantics but not \textit{ambiguity blocking}.

A Dung-like version of \textit{ambiguity blocking} has been sketched in the context of instantiating defeasible logics semantics and the Carneades argumentation systems into ASPIC+ \cite{van2012relating}, a structured argumentation framework based on Dung’s semantics \cite{modgil2014aspic+}.
Instead of defining a new Dunganian semantics, the authors proposed a solution based on the introduction of additional arguments unidirectionally attacking the conflicting ambiguous arguments to reject both of them and \textit{replicate} the \textit{ambiguity blocking} behaviour. In \cite{lam2016aspic+} the author noted how this solution requires to introduce a second “attack” relation on arguments with a ripple down effects on the ASPIC+ definitions setting the various statuses of the argument.

Our approach differs substantially from previous attempts. Our aim is to introduce a new abstract semantics where the mechanism of \textit{ambiguity blocking} is embedded in the postulates of the semantics rather than being mimicked by additional arguments. 
Our solution rely on connecting the notion of \textit{ambiguity blocking} with \textit{undecidedness blocking}. Indeed the concepts of \textit{undecidedness blocking} and \textit{ambiguity blocking} are similar but not equivalent and we will show how undecidedness encompasses a larger variety of situations besides ambiguity.

\vspace{2mm}
\noindent \textit{Motivation 2: The weak attacks problem: do arguments have to be defended against all the attacks?}
\vspace{2mm}

The problem is illustrated in graph $B$ of figure 1. It is the case of a self-defeating argument $a$ attacking argument $b$. 
Recently, this example was analysed by Baumann et al. \cite{bbu1} \cite{baumann2020comparing}, leading to the definition of novel abstract semantics based on a weaker notion of admissibility, referred here as $BBU$ semantics.
Dung already stated the problem (\cite{dung} p. 351):
\textit{`An interesting topic of research is the problem of self-defeating arguments as illustrated in the following example (figure 1B). The only preferred extension here is empty though one can argue that since A defeats itself, B should be acceptable'}.\\
The observation is extended to situations like the one in graph $C$ in figure 1 where the attacker is not self-defeating but nevertheless it is part of an odd-length cycle of arguments collectively defeating themselves. 
The problem of self-defeating attacking arguments described so far is indeed strictly related to the problem of the propagation of undecidedness described in motivation 1. The general questions are the following: 

\textit{Do arguments have to be defended against all attackers? When can we consider an attack weak enough to be neglected (blocked) instead of being considered effective (propagated)?}

Those questions are rephrasing the concept of \textit{ambiguity propagation} vs. \textit{ambiguity blocking} present in other non-monotonic formalisms. In legal reasoning, the questions are answered by the different standards of proof.

Clearly, both the semantics proposed in this paper and the $BBU$ semantics answer the above questions and they are both non-admissible semantics. A comparison is therefore important and it is detailed in section 6.3.
We first claim how our semantics are the first to tackle the above problems. Indeed, the semantics presented in this paper expand our previous works in \cite{dondio2019beyond} \cite{dondio2018proposal} that chronologically precede $BBU$ semantics by more than one year. In \cite{dondio2019beyond} \cite{dondio2018proposal} we introduced new semantics \textit{"based on a weaker notion of admissibility but retaining the majority of principles of complete semantics"} (\cite{dondio2018proposal}, pg. 4) to answer the above questions. We wondered what attacks are weak enough to be neglected. Our work moved from legal reasoning and standards of proof to the concept of undecidedness as a form of weaker attack, concept that we fully develop in this work.
However in \cite{dondio2019beyond} some of the semantics presented here are already defined. Those semantics are able to handle correctly all examples of figure 1, therefore including also $BBU$’s motivating examples.
In \cite{dondio_tech} we also sketched the \textit{weakly complete} semantics that are developed in this work.
Therefore, we claim that our work is indeed independent and prior to $BBU$ semantics. In particular, the problem of \textit{weak attacks} was first clearly answered by the authors of this paper.

Our answer to \textit{'which arguments are weak enough to be neglected'} is nevertheless different from $BBU$’s one.
Both of the solutions remove the admissibility property of Dung’ semantics.
However, the semantics do not coincide. Our \textit{undecidedness blocking} semantics are able to handle all the motivating examples of figure 1, while $BBU$ semantics do not address the problem of graph A, figure 1. 
The reader could object that graph $A$ is not in the list of $BBU$ motivating examples. However, our semantics allow for an interpretation by which graphs $A$ and $B$ of figure 1 are variations of the same situation: argument $b$ is attacked by a self-conflicting argument in graph $B$, and it is attacked by two conflicting arguments in graph $A$. If we take a \textit{grounded} semantics stance on the two rebutting arguments $a$ and $c$ of graph $A$, we could make the case that $a$ and $c$ are also self-defeating and $b$ should be accepted, as it happens with \textit{ambiguity blocking} semantics. 

Substantial differences between the two families of semantics are revealed by a principle-based comparison detailed in section 6.3. The analysis shows how our semantics satisfy a number of principles satisfied by Dung's \textit{complete} semantics but not by Baumann et al. semantics, including \textit{directionality}, \textit{abstention}, \textit{SCC-decomposability} and \textit{cardinality} of extensions, making them a more faithful not admissible version of Dung’ semantics.

\vspace{2mm}

The paper is organized as follows.
Section 2 provides the required background knowledge of abstract argumentation while section 3 is dedicated to the presentation of \textit{weakly complete} semantics. The presentation of the new semantics is self-contained and it does not refer to \textit{ambiguity blocking} mechanisms present in other formalisms. In Section 4 we present an algorithm to compute \textit{weakly complete} semantics while in section 5 we introduce several families of \textit{weakly complete} semantics. Section 6 presents a principle-based analysis of our semantics and a comparison with recent approaches. Section 7 contains related works to date and section 8 our future works and conclusions.

\section{Abstract argumentation semantics} \label{sec:abstractArg}
In this section we provide the concepts of abstract argumentation semantics required for the remaining of this paper.\\

\noindent \textbf{Definition 2.1} An \textit{argumentation framework  $AF$ is a pair $\langle Ar,\mathcal{R} \rangle$, where $Ar$ is a non-empty finite set whose elements are called arguments and $ \mathcal{R} \subseteq Ar \times Ar $ is a binary relation, called the attack relation. If $(a,b) \in \mathcal{R} $ we say that $a$ attacks $b$. An argument is \textit{initial} if it is not attacked by any arguments, including itself}.\\

An abstract argumentation semantics identifies the sets of arguments that can survive the conflicts encoded by the attack relation $\mathcal{R}$. Dung's semantics \cite{dung} require a group of acceptable arguments to be conflict-free (an argument and its attackers cannot be accepted together) and admissible (the set of acceptable arguments has to defend itself from external attacks). \\

\noindent  \textbf{Definition 2.2} (conflict-free set). \textit{A set $Arg \subseteq Ar$ is conflict-free iff $\forall a,b \in Arg, (a,b) \not\in \mathcal{R}$}.\\

\noindent \textbf{Definition 2.3} (admissible set, complete set). \textit{A set $Arg \subseteq Ar$ defends an argument $a \in Ar$ iff $\forall b \in Ar $ such that $(b,a) \in \mathcal{R} , \exists c \in Arg$ such that $(c,b) \in \mathcal{R}$. The set of arguments defended by $Arg$ is denoted $\mathcal{F}(Arg)$. A conflict-free set $Arg$ is \textit{admissible} if $Arg \subseteq\mathcal{F}(Arg)$ and it is \textit{complete} if $Arg=\mathcal{F}(Arg)$}\\

In this paper we follow the labelling approach of \cite{caminada}, where a semantics assigns to each argument a label \texttt{in}, \texttt{out} or \texttt{undec}. \\

\noindent \textbf{Definition 2.4} (labelling). \textit{Let $AF=\langle Ar,\mathcal{R} \rangle$. A labelling is a total function} $\mathcal{L}: Ar \to \{\texttt{in},\texttt{out},\texttt{undec\}}$. \textit{We write $in(\mathcal{L})$ for $\{a\in Ar | \mathcal{L}(a)=\texttt{in}\}$, $out(\mathcal{L})$ for $\{a\in Ar | \mathcal{L}(a)=\texttt{out}\}$, and $undec(\mathcal{L})$ for $\{a\in Ar | \mathcal{L}(a)=\texttt{undec}\}$}. \\

\noindent \textbf{Definition 2.5} (from \cite{caminada}). \textit{Let $AF=\langle Ar,\mathcal{R} \rangle$. A \textbf{complete labelling} is a labelling such that for every $a \in Ar$ holds that}:
\begin{enumerate}
\renewcommand\labelenumi{\theenumi}
\renewcommand{\theenumi}{\arabic{enumi}.}
\item \textit{if $a$ is labelled} \texttt{in} \textit{then all its attackers are labelled} \texttt{out};
\item \textit{if $a$ is labelled} \texttt{out} \textit{then it has at least one attacker that is labelled} \texttt{in}; 
\item \textit{if $a$ is labelled} \texttt{undec} \textit{then it has at least one attacker labelled} \texttt{undec} \textit{and it does not have an attacker that is labelled} \texttt{in}.\\
\end{enumerate}

\noindent \textbf{Definition 2.6} (complete labelling \cite{caminada}) \textit{Given $AF=(Ar,\mathcal{R})$, $\mathcal{L}$ is the \textit{grounded} labelling iff $\mathcal{L}$ is a complete labelling where $undec(\mathcal{L})$ is maximal (w.r.t. set inclusion) among all complete labellings of $AF$. $\mathcal{L}$ is the \textit{preferred} labelling iff $\mathcal{L}$ is a complete labelling where $in(\mathcal{L})$ is maximal (w.r.t. set inclusion) among all complete labellings of $AF$. A labelling $\mathcal{L}$ is \textit{stable} iff $undec(\mathcal{L}) = \emptyset $.
}\\

Referring to Figure 2, there is only one complete labelling for $G_{1}$, representing also the \textit{grounded, preferred} and \textit{stable} labelling, where argument $a$ is \texttt{in} (no attackers), $b$ is \texttt{out} and $c$ is \texttt{in}. 
Regarding $G_{2}$, \textit{grounded} semantics assign the \texttt{undec} label to all the arguments. Regarding the preferred semantics, there are two complete labellings that maximise the $in(\mathcal{L})$ set: one with $in(\mathcal{L}_{1})=\{b,d\}$, $out(\mathcal{L}_{1})=\{c,a,e\}$, $undec(\mathcal{L}_{1})=\emptyset$ and the other with $in(\mathcal{L}_{2})=\{a\}$, $out(\mathcal{L}_{2})=\{b\}$, $undec(\mathcal{L}_{2})=\{c,d,e\}$. 
\begin{figure}[h]
\begin{center}
\includegraphics[scale=0.6]{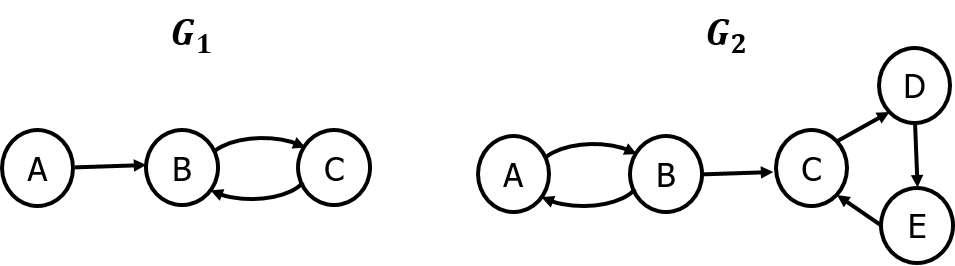}
\caption{Two Argumentation graphs}
\label{fig1}

\end{center}
\end{figure}

Besides the labelling approach, one can follow an extension-based approach, where a semantics identifies the sets of accepted argument, called extensions. 
The labelling and the extension-based approach are equivalent: given an argumentation framework, for each extension identified by a semantics there is an equivalent set of \texttt{in}-labelled arguments and vice versa (see \cite{caminada}).
An argumentation framework $AF=\langle Ar,\mathcal{R} \rangle$ identifies a directed graph. The following are some graph-based definitions needed in this paper.\\

\noindent \textbf{Definition 2.7} \textit{A (vertex-induced) subgraph of a graph $G=(Ar,\mathcal{R})$ is a graph $G_s=(S,\mathcal{R}_s)$, where $S \subseteq Ar$ and $\mathcal{R}_s=\mathcal{R} \cap (S \times S)$}.\\

\noindent A subgraph contains a subset of nodes of the original graph and any link whose endpoints are both in $S$ (note how this subgraph is usually called a \textit{vertex-induced} subgraph). Given an argumentation framework $AF$, the restriction of an argumentation framework to a set of nodes $S$ is the framework $AF_{\downarrow S}$ corresponding to the vertex-induced subgraph of $AF$ identified by the nodes $S$. \\

\noindent \textbf{Definition 2.8} \textit{If $G$ is a graph, a \textit{strongly connected graph} of $G$ is a subgraph of $G$ where, for each pair of nodes $a,b \in G$ there is at least one directed path from $a$ to $b$ and at least one directed path from $b$ to $a$. A \textit{strongly connected component} (SCC) of $G$ is a \textit{maximal} (with respect to set inclusion) strongly connected graph of $G$}.\\

\noindent A \textbf{cycle} in a graph is a non-empty directed path in which the only repeated vertices are the first and last. Note how a node that is not part of a cycle is a strongly connected component. In this paper we call \textbf{acyclic argument} an argument that is not part of a cycle, and \textbf{cyclic argument} otherwise. Self-attacking arguments are cyclic arguments. We also refer to \textbf{cyclic strongly connected component} to identify a strongly connected component containing at least one cycle.

\section{Weakly complete semantics}\label{sec:weaklySemantics}
In this section \textit{weakly complete} semantics are formally defined. Our aim is to define a semantics whose postulates control the propagation of the undecided label. We have already seen how in all Dung's \textit{complete} semantics the label \texttt{undec} is always propagated from the attacker to the attacked argument, unless the former is attacked by an accepted argument as well. Anyhow the attacked argument is never accepted. 
\textit{Weakly complete} semantics employ a mechanism to \textit{block} the propagation of the undecided label, so that an argument can be accepted even if attacked by undecided arguments. The main consequence is the loss of the \textit{admissibility} property.
As in \textit{complete} semantics, an argument is rejected if and only if it has at least one \texttt{in}-labelled attacker. The definition of \textit{weakly complete} semantics requires only a small modification of definition 2.5.\\

\noindent \textbf{Definition 3.1} \textit{Given an argumentation framework $AF = \langle Ar,\mathcal{R} \rangle$, a \textit{weakly complete} labelling is one such that for every $a \in Ar$ it holds that}:
\begin{enumerate}
\renewcommand\labelenumi{\theenumi}
\renewcommand{\theenumi}{\arabic{enumi}.}
\item \textit{if $a$ is labelled} \texttt{in} \textit{then there are no attackers of $a$ labelled} \texttt{in};
\item \textit{if $a$ is labelled} \texttt{out} \textit{then it has at least one attacker that is labelled} \texttt{in}; 
\item \textit{if $a$ is labelled} \texttt{undec} \textit{then it has at least one attacker labelled} \texttt{undec} \textit{and it does not have an attacker that is labelled} \texttt{in}.
\end{enumerate}

The above definition changes condition 1 of definition 2.5 of Dung's \textit{complete} labelling by relaxing it, since now there is the option to accept an argument attacked by undecided arguments. These arguments would be always labelled \texttt{undec} in a complete labelling. 
In Table 1 we have broken down and named the conditions included in definition 3.1.

\begin{table}[h]
\centering
\begin{tabular}{|c|p{35mm} | p{65mm}|}
\hline
 &\textbf{Condition Name} & \textbf{Description} \\
\hline
\textit{3.1a}&\textit{Admissiblity} \textbf{(ad)} & $a$ is \texttt{in} if all its attackers are \texttt{out}  \\
\hline
\textit{3.1b}&\textit{Undecidedness Blocking} \textbf{(ub)}& $a$ is \texttt{in} if at least one attacker is \texttt{undec} and all the others are \texttt{out} \\
\hline
\textit{3.1c}&\textit{Undecidedness Propagating} \textbf{(up)}& $a$ is \texttt{undec} if at least one attacker is \texttt{undec} and all the others are \texttt{out} \\
\hline
\textit{3.1d}&\textit{Rejection} \textbf{(re)}& $a$ is \texttt{out} if at least one attacker is \texttt{in}  \\
\hline\end{tabular}
\caption{Conditions for \textit{weakly complete} semantics}
\end{table}

\noindent \textit{Weakly complete} labellings generated without using the \textit{undecidedness blocking} condition \textit{3.1b} are \textit{complete}, and therefore all \textit{complete} labellings are \textit{weakly complete}. Labellings whose at least one label satisfies condition \textit{3.1b} instead of condition \textit{3.1c} are non-admissible.

\subsection{Examples of Weakly Complete Labellings}
We present some examples of \textit{weakly complete} labellings to familiarize with the new semantics introduced.
Figure 3 displays five argumentation frameworks and their \textit{weakly complete} labellings. Note how the first three graphs are also included in the motivating examples of figure 1.
In the graph $G_{3}$ (a cycle of three arguments) the only valid \textit{weakly complete} labelling is the \textit{grounded} one where all the arguments are undecided.
In $G_{4}$ there is an additional labelling besides the \textit{grounded} one, where argument $b$ is accepted by blocking the undecidedness from $a$ to $b$. 
$G_{5}$ is an example of \textit{floating assignment} graph. There are four \textit{weakly complete} labellings. Besides the \textit{grounded} labelling (all arguments are labelled undecided) there are two additional \textit{preferred} labellings, in both of which argument $c$ is rejected. 
There is a new additional \textit{weakly complete} labelling, where $a$ and $b$ are undecided and $c$ is accepted, therefore handling in the expected way our motivating example. 
$G_{6}$ will be discussed in details in the next section.
$G_{7}$ shows an example of an argumentation graph consisting of a single strongly connected component. Besides the \textit{grounded} labelling, there is also a non-admissible \textit{weakly complete} one, in which the cycle $a,b,c$ is a source of unresolved conflict and labelled \texttt{undec}, but the \textit{undecidedness blocking} condition is applied to the attack from $c$ to $d$ to generate a valid labelling where $d$ is accepted and $e$ rejected.

\begin{figure}[h]
\begin{center}
\includegraphics[scale=0.40]{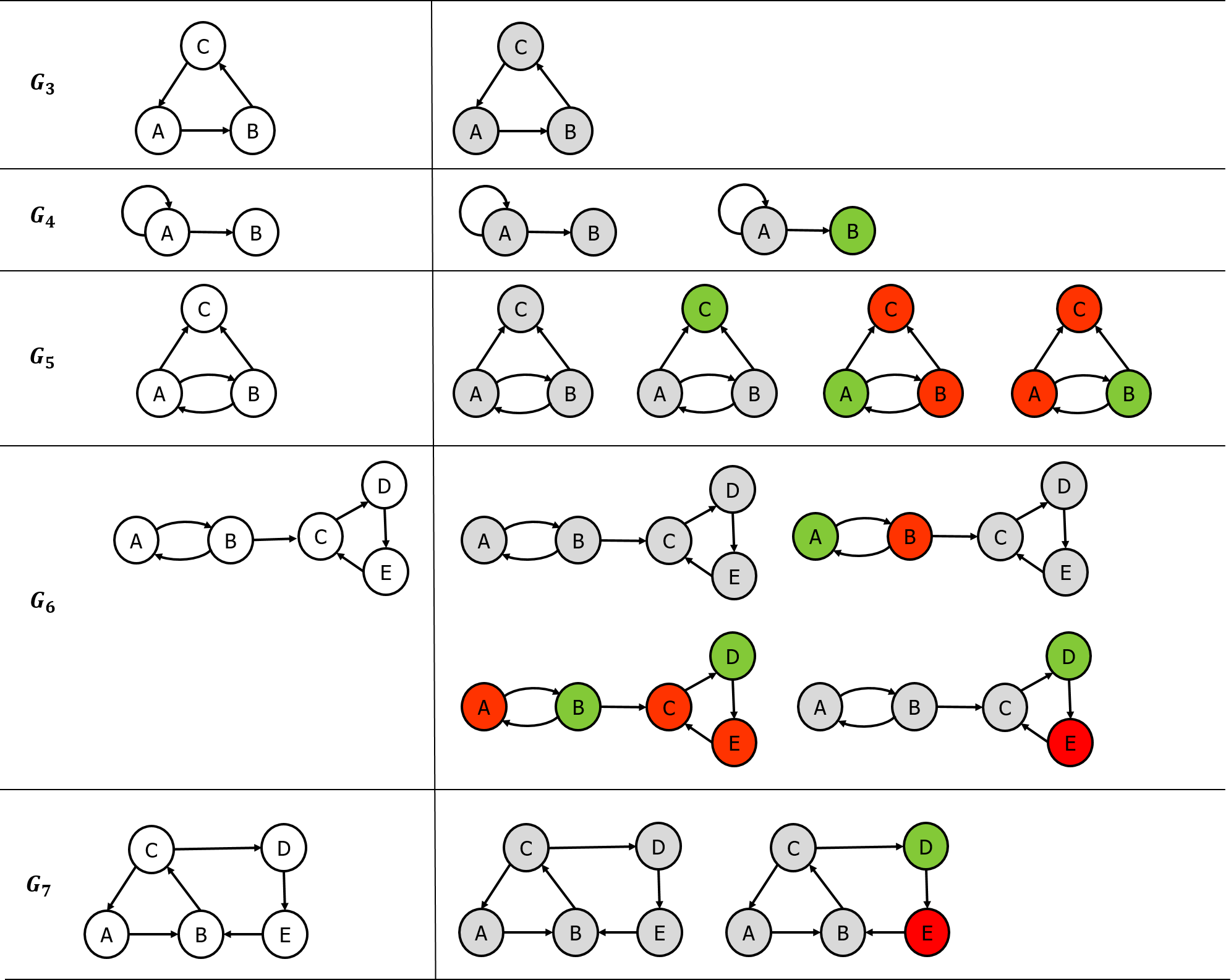}
\caption{Examples of \textit{weakly complete} labellings}
\label{fig1}
\end{center}
\end{figure}

We already observed that \textit{weakly complete} labellings generated without using the \textit{undecidedness blocking} condition \textit{3.1b} are \textit{complete}, while \textit{weakly complete} labelling using condition \textit{3.1b} instead of condition \textit{3.1c} are non-admissible. In general there are \textit{weakly complete} labelling where a combination of conditions \textit{3.1b} and \textit{3.1c} is used. These are the labellings where the undecided label is propagated to some parts of the argumentation graph but not to all of them. These labellings represent interesting cases, where an agent might grant \textit{undecidedness blocking} to some arguments, preventing them to be labelled undecided, but not to others. An example is shown in Figure 4. The figure shows three labellings of the same argumentation framework. The first is the \textit{grounded} labelling where the undecided label is always propagated, the other two are \textit{weakly complete} ones. In the labelling on the centre the \textit{ub}-condition \textit{3.1b} is applied \textit{earlier} to the attack from $b$ to $d$, while in the labelling on the right the condition is applied to the attack from $d$ to $e$, but not to the attack from $b$ to $d$, where the \textit{undecidedness propagating} condition \textit{3.1c} is used instead. 

\begin{figure}[h]
\begin{center}
\includegraphics[scale=0.48]{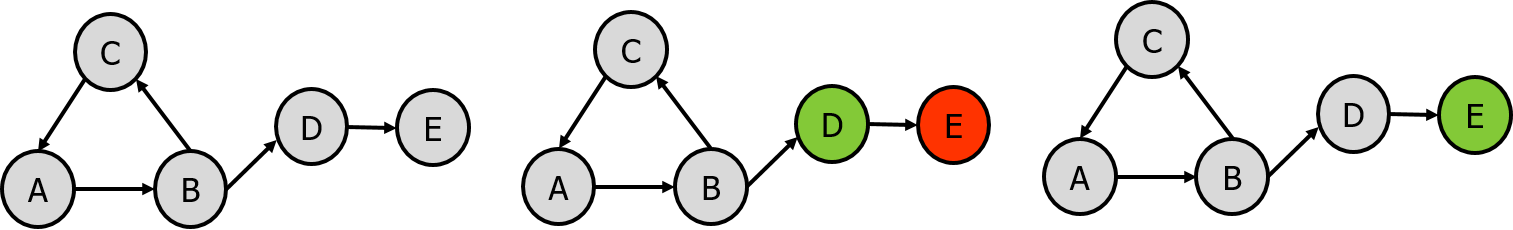}
\caption{Three labellings where undecidedness is propagated (left), blocked at argument $d$ (center) or at argument $e$ (right).}
\label{fig1}
\end{center}
\end{figure}

\section{Computing Weakly Complete Semantics}\label{sec:algoWeaklySemantics}
In this section we propose an algorithm to compute \textit{weakly complete} semantics, firstly introducing some preliminary notations. Given an argumentation framework $AF = \langle Ar,\mathcal{R} \rangle$, we call $\{a\}^-=\{x \in Ar | \; (x,a) \in \mathcal{R}\}$ the set of arguments in $AF$ attacking argument $a$ and $\{a\}^+=\{x \in Ar | \; (a,x) \in \mathcal{R}\}$ the set of arguments in $AF$ attacked by argument $a$.
We start by observing that, as for \textit{complete} semantics, a \textit{weakly complete} labelling $\mathcal{L}$ is fully identified by the set $in(\mathcal{L})$.\\

\noindent \textbf{Proposition 4.1} \textit{Let us consider $AF = \langle Ar,\mathcal{R} \rangle$ and a weakly complete labelling $\mathcal{L}$ of $AF$. Then, the sets $out(\mathcal{L})$ and $undec(\mathcal{L})$ are uniquely identified by $in(\mathcal{L})$}. It is $out(\mathcal{L})=\{ a \in Ar | \exists b \in Ar \land \mathcal{R}(b,a) \land \mathcal{L}(b)=$ \texttt{in}$\}$. \\

\noindent \textit{Proof}.
We label all the arguments of $AF$ with a label \texttt{unk} (=\textit{unknown}). Then, we assign the label \texttt{in} to the arguments in $in(\mathcal{L})$ and for the \textit{rejection} condition \textit{3.1d} we necessarily assign the label \texttt{out} to the arguments attacked by at least one argument in $in(\mathcal{L})$. Since $in(\mathcal{L})$ is conflict-free, none of the arguments in $in(\mathcal{L})$ is changed to the label \texttt{out}. The remaining \texttt{unk} arguments cannot be labelled \texttt{out} since they do not have any attacker in $in(\mathcal{L})$ and they have at least one attacker labelled \texttt{unk}, otherwise they would be labelled \texttt{in} and therefore part of $in(\mathcal{L})$. Therefore the label \texttt{undec} is the only label that can be assigned to the \texttt{unk} arguments. We have to verify that the resulting labelling is valid.
Since \texttt{unk} arguments are attacked by at least one \texttt{unk} argument and arguments labelled \texttt{out}, by changing the \texttt{unk} label to \texttt{undec}, each undecided argument results attacked by at least one undecided argument (previously \texttt{unk}) and by \texttt{out}-labelled arguments, and therefore the undecided labels are valid for the condition \textit{3.1d} of \textit{weakly complete} semantics. Regarding the arguments labelled \texttt{in}, their attackers are either \texttt{out} or \texttt{undec}, and therefore the labels \texttt{in} are valid in a  \textit{weakly complete} labelling for the \textit{undecidedness blocking} condition \textit{3.1b}. $\square$ \\

Under both \textit{complete} and \textit{weakly complete} semantics, assigning the label \texttt{in} to an argument $a$ imposes constraints on the label of other connected arguments. It imposes that all the arguments $\{a\}^+$ are labelled \texttt{out}. As a consequence, if an argument $b$ has now all its attackers in $\{a\}^+$ it can be labelled \texttt{in}. In turn, this imposes to all the arguments in $\{b\}^+$ the label \texttt{out} and so forth, stopping when no more labels can necessarily be imposed on arguments or an inconsistency is found when trying to label the same argument twice with different labels.
The reasoning can be extended to a set of arguments $Gr \subseteq Ar$ instead of a single argument. Assigning the label \texttt{in} to all the arguments in $Gr$ imposes similar constraints.
This \textit{forward propagation} of the \texttt{in} and \texttt{out} labels is formalized in the \texttt{in-out-fw} algorithm.

\begin{algorithm}[h] \label{algor}
\footnotesize
\SetAlgoLined
\textbf{Inputs}: {$AF = \langle Ar,\mathcal{R} \rangle$, $\mathcal{L}$ is a labelling of $AF$, $Gr \subseteq undec(\mathcal{L}) \subseteq Ar$, }\;
\textbf{Outputs}: {an updated labelling $\mathcal{L}$ or inconsistency}\;
\For{g $\in Gr$} {
$\mathcal{L}(g) \leftarrow$ \texttt{in}\;
}
\Repeat{$Gr$ = $\emptyset$} {
$g \leftarrow $ \textbf{pop}(Gr)\;
$\mathcal{L}(g) \leftarrow$ \texttt{in}\;
\For{x $\in \{g\}^+$}{
\eIf {$\mathcal{L}(x)$=\texttt{in}}  {\textbf{return inconsistency}\;
} {$\mathcal{L}(x) \leftarrow$ \texttt{out}\;}
}
$Gr \leftarrow \{x \in Ar \: | \: \mathcal{L}(x) =\texttt{undec} \land \forall a \in \{x\}^- \;  \mathcal{L}(a)=$\texttt{out}  $\}$ \;
}
\textbf{return} $\mathcal{L}$
 \caption{The \texttt{in-out-fw} algorithm}
\end{algorithm}

The inputs of the \texttt{in-out-fw} algorithm are an argumentation framework $AF$, a set of arguments $Gr \subseteq Ar$ and a labelling $\mathcal{L}$ of $AF$.
The output is either an inconsistency error or an updated version of $\mathcal{L}$. The algorithm first assigns the label \texttt{in} to all the  arguments $a \in Gr$. Then, for each arguments $a$ in $Gr$, the label \texttt{out} is assigned to the arguments in $\{a\}^{+}$ and, as a consequence, the arguments those attackers are all labelled \texttt{out} are added to the set $Gr$ of new arguments that could be promoted to the label \texttt{in}. A new argument in $Gr$ is selected and the labelling process is repeated until $Gr$ is empty or we found an inconsistency. In case of inconsistency all the modifications to $\mathcal{L}$ are discarded.

Given a labelling $\mathcal{L}$ of $AF$, we refer to the application of the algorithm using the set of argument $Gr$ as \texttt{in-out-fw}$(Gr,\mathcal{L}$). Each argument in $Gr$ is called a \textbf{\textit{ground}} of the algorithm. 

Since the algorithm uses only the admissibility (\textit{3.1a}) and rejection (\textit{3.1d}) conditions, all the labels assigned by the \texttt{in-out-fw}  algorithm (excluding the one assigned to the ground argument) are valid for both \textit{complete} and \textit{weakly complete} semantics.
Moreover all the labels assigned by the \texttt{in-out-fw}$(Gr,\mathcal{L}$) algorithm follow necessarily form accepting the \textit{ground} arguments in $Gr$ and the labels assigned in $\mathcal{L}$. If an inconsistency is found, this means that there is no \textit{weakly complete} labelling where the arguments in $Gr$ are accepted together with the arguments in $in(\mathcal{L})$; at least one argument in $Gr$ or in $in(\mathcal{L})$ cannot be accepted. On the contrary, if $\mathcal{L}^{'}=$ \texttt{in-out-fw}$(Gr,\mathcal{L}$) ends without an inconsistency, if $\mathcal{L}$ was not a \textit{complete} or \textit{weakly complete} labelling we are not guaranteed that $\mathcal{L}^{'}$ is \textit{complete} or \textit{weakly complete}. 

In this section we show how any \textit{weakly complete} labelling can be obtained by repeatedly applying the \texttt{in-out-fw} algorithm. Therefore, we start by providing some properties of the labellings generated by the \texttt{in-out-fw} algorithm needed to prove the main results of this section.  
\noindent Given $AF = \langle Ar,\mathcal{R} \rangle$, we call $\mathcal{L}_{und}$ the labelling function assigning to each argument in $Ar$ the label \texttt{undec}. 
The following is a corollary of Proposition 4.1:\\

\noindent \textbf{Corollary 4.1} \textit{Given $AF = \langle Ar,\mathcal{R} \rangle$ and a weakly complete labelling $\mathcal{W}$ of $AF$, it is $\mathcal{W}=$}  \texttt{in-out-fw}$(in(\mathcal{W}),\mathcal{L}_{und})$.\\

\noindent \textit{Proof}.
The algorithm \texttt{in-out-fw}$(in(\mathcal{W}),\mathcal{L}_{und})$ assigns the label \texttt{in} to all the arguments in $in(\mathcal{W})$ and \texttt{out} to all the arguments attacked by an argument in $in(\mathcal{W})$. Since $\mathcal{W}$ is conflict-free, none of the arguments in $in(\mathcal{W})$ are labelled \texttt{out}. Then the algorithm stops since the remaining undecided arguments are at least attacked by an undecided argument. Indeed, if an undecided argument $a$ is attacked only by \texttt{out}-labelled arguments, $a$ should necessarily be in $in(\mathcal{W})$, and therefore $\mathcal{W}$ is not a valid \textit{weakly complete} labelling since if $a$ is \texttt{undec} or \texttt{out}, the admissibility property would be violated and $\mathcal{W}$ would not be a \textit{weakly complete} labelling either.
By Proposition 4.1, the labelling found is $\mathcal{W}$.   $\square$ \\

\noindent The following straightforward property is useful in the remaining discussion.\\

\noindent \textbf{Proposition 4.2} \textit{Each application of the} \texttt{in-out-fw} \textit{algorithm does not change the labels of arguments previously labelled} \texttt{in} \textit{or} \texttt{out}.\\

\noindent \textit{Proof}.
The \texttt{in-out-fw} algorithm cannot change a label of an argument previously labelled \texttt{in}, since that would generate an inconsistency. Neither it cannot change the label of an argument $a$ previously labelled \texttt{out}, since that would require to change the label \texttt{in} of one or more attackers of $a$. $\square$ \\

An interesting property of the  \texttt{in-out-fw} algorithm is that it can be applied to a set of arguments $Gr$ or sequentially to each argument in $Gr$ and the final labelling is the same, since the algorithm generates the same set of constrains on the labels of the arguments in $AF$. The order of the arguments in $Gr$ is also irrelevant. Indeed, each consistent application of the \texttt{in-out-fw} algorithm does not change the previously labelled arguments, and arguments requiring constraints from more than one \textit{ground} arguments in order to be labelled are labelled only when the last of the required \textit{ground} argument's \texttt{in-out-fw} is executed, independently from the order of execution. For example,  an  argument  defended  by the set of arguments $\{a,b\}$ can  be  labelled \texttt{in} only  when both \texttt{in-out-fw}$(a)$ and \texttt{in-out-fw}$(b)$  are  executed. We formally express this property of the  \texttt{in-out-fw} algorithm: \\

\noindent \textbf{Property 4.1} \textit{Given a framework $AF = \langle Ar,\mathcal{R} \rangle$ and its labellings $\mathcal{L}_{ab}=$} \texttt{in-out-fw}$(\{a,b\},\mathcal{L}_{und})$, $\mathcal{L}_{a}=$ \texttt{in-out-fw}$(\{a\},\mathcal{L}_{und})$,  $\mathcal{L}_{b}=$ \texttt{in-out-fw}$(\{b\},\mathcal{L}_{und})$. \textit{If} $\mathcal{L}_{ab} \not = \mathcal{L}_{und}$ \textit{then} $\mathcal{L}_{ab}=\mathcal{L}_{a}=\mathcal{L}_{b}$.\\

\noindent The above proposition is true only if \texttt{in-out-fw}$(\{a,b\},\mathcal{L}_{und})$ ends without an inconsistency ($\mathcal{L}_{ab} \not = \mathcal{L}_{und}$). For instance if we consider the framework $G_5$ and the set $\{a,c\}$,  it is \texttt{in-out-fw}$(\{a,c\},\mathcal{L}_{und})=\mathcal{L}_{und}$ but if we apply \texttt{in-out-fw} frist to $a$ and then to $c$ we obtain the labelling $\{\{a\},\emptyset,\{b,c\}\}$ and if we apply \texttt{in-out-fw} first to $c$ and then to $a$ we obtain the labelling $\{\{c\},\emptyset,\{a,b\}\}$.

The following proposition shows that if 
$\mathcal{L}=$ \texttt{in-out-fw}$(A,\mathcal{L}_{und})$ does not generate an inconsistency, then all the arguments accepted in $\mathcal{L}$ are also accepted in any \textit{weakly complete} labelling accepting the set of arguments $A$.\\

\noindent \textbf{Proposition 4.3} \textit{Given a framework $AF = \langle Ar,\mathcal{R} \rangle$, let us consider the labelling of $AF$ $\mathcal{L}=$}\texttt{in-out-fw}$(A,\mathcal{L}_{und})$. \textit{If $\mathcal{W}$ is a weakly complete labelling with $A \subseteq in(\mathcal{W)}$, then $in(\mathcal{L)} \subseteq in(\mathcal{W)}$.}\\

\noindent \textit{Proof}.
According to property 4.1, we can \textit{build} the labelling $\mathcal{W}$ sequentially starting from $\mathcal{L}_{und}$ and using all the arguments in $in(\mathcal{W})$. Starting from $\mathcal{L}_{und}$, we applying the \texttt{in-out-fw} algorithm to all the arguments in $A$, obtaining $\mathcal{L}$.
We then apply the \texttt{in-out-fw} algorithm to all the arguments in $in(\mathcal{W}) \setminus A$ to obtain $\mathcal{W}$. None of the applications of the \texttt{in-out-fw} can generate an inconsistency (otherwise the \textit{ground} argument generating the inconsistency would not be in $in(\mathcal{W})$, a contradiction) and since no application of the \texttt{in-out-fw} algorithm can change the labels \texttt{in} and \texttt{out} previously assigned (Proposition 4.2), it is $in(\mathcal{L)} \subseteq in(\mathcal{W)}$. $\square$ \\

\indent We can prove the following proposition useful in the remaining discussion:\\

\noindent \textbf{Proposition 4.4} \textit{Given $AF = \langle Ar,\mathcal{R} \rangle$, let us consider its labelling $\mathcal{L}_{gr}=$} \texttt{in-out-fw}$(Gr,\mathcal{L}_{und})$. \textit{We consider $g \in Ar \setminus Gr$. If} \texttt{in-out-fw}$(\{g\},\mathcal{L}_{gr})$ \textit{generates an inconsistency, the inconsistency is generated when trying to label} \texttt{out} \textit{the argument $g$ or an argument in the set of ground $Gr$}.\\

\noindent \textit{Proof}.
We observe that in $\mathcal{L}_{gr}$ the arguments previously used as ground (set $Gr$) that did not generate an inconsistency are labelled \texttt{in} and they might have some attacking arguments labelled \texttt{undec}. Differently, the labels of the arguments not in $Gr$ follows the \textit{admissibility} or \textit{rejection} condition: if they are labelled \texttt{out} they have at least one attacker labelled \texttt{in} and if they are labelled \texttt{in} they have all their attackers labelled \texttt{out}. 

We prove that an inconsistency cannot be generated when the \texttt{in-out-fw} algorithm  labels an argument $a \not \in Gr \cup g$.
When the ground $g$ is promoted from \texttt{undec} to \texttt{in}, arguments in $\{g\}^+$ are changed to the label \texttt{out}. An inconsistency is generated only if an argument $a$ in $\{g\}^+$ had $\mathcal{L}_{gr}(a)=$ \texttt{in}. However, if an argument $a$ in $\{g\}^+$ is not in $Gr \cup g$, an inconsistency cannot be generated since  $\mathcal{L}_{gr}(a) \not =$ \texttt{in}; since $a \in \{g\}^+$ and therefore $a$ is attacked by $g$ that was undecided in $\mathcal{L}_{gr}$. Argument $a$ was either labelled \texttt{out}  because of the effect of an \texttt{in}-labelled attacker distinct from $g$ or labelled \texttt{undec}  otherwise. Therefore no inconsistency can be generated when trying to label arguments not in $Gr \cup g$.

The effect of labelling \texttt{out}  the arguments in $\{g\}^+$ could be to promote some arguments in $\{g\}^{++}=\bigcup \{h\}^+ , h \in \{g\}^+$ to the label \texttt{in}. 
Those arguments were labelled \texttt{in}  or \texttt{undec} in $\mathcal{L}_{gr}$ but not \texttt{out}, since arguments in $\{g\}^{++}$ are attacked by arguments in $\{g\}^+$ that were not labelled \texttt{in} in $\mathcal{L}_{gr}$ as shown above.

Changing the label of an argument $a \in \{g\}^{++}$ to \texttt{in} generates the same constraints we discussed when the label \texttt{in} was assigned to the ground argument $g$, concluding that if at each step the attacked arguments $\{a\}^{+}$ are not in $Gr \cup g$, no inconsistency can be generated. We also observe that only the labels of arguments undecided in $\mathcal{L}_{gr}$ are changed by the effect of \texttt{in-out-fw} algorithm.

Therefore, the only way to generate an inconsistency is to have an argument $x$ with $\mathcal{L}_{gr}(x)=$ \texttt{in}  that had at least one undecided attacker in $\mathcal{L}_{gr}$ that is changed to the label \texttt{in} by  \texttt{in-out-fw}$(g,\mathcal{L}_{gr})$. Since $\mathcal{L}_{gr}(x)=$ \texttt{in}  and $x$ had at least an undecided attacking argument in $\mathcal{L}_{gr}$, $x$ must be in $Gr \cup g$ and therefore the thesis is proven. $\square$ \\

\noindent Given $AF = \langle Ar,\mathcal{R} \rangle$, we call $I(AF)=\{a \in Ar | \nexists b \in Ar \land \mathcal{R}(b,a) \}$ the set of initial arguments of $AF$. We consider the labelling $\mathcal{L}_{I}$ of $AF$ obtained by \texttt{in-out-fw}$(I(AF),\mathcal{L}_{und})$. \\
\noindent According to Proposition 4.4, \texttt{in-out-fw}$(I(AF), \mathcal{L}_{und})$ cannot generate an inconsistency, since the ground arguments are in $I(AF)$ and therefore they cannot be labelled twice by the \texttt{in-out-fw} algorithm. Therefore $I(AF) \subseteq$ \textit{in($\mathcal{L}_{I}$)}. We prove that $\mathcal{L}_{I}$ is the \textit{grounded} labelling of $AF$.\\

\noindent \textbf{Proposition 4.5} \textit{Let us consider $AF = \langle Ar,\mathcal{R} \rangle$ and its \textit{grounded} labelling $\mathcal{G}_{AF}$. Then $\mathcal{L}_{I} = \mathcal{G}_{AF}$}.\\

\noindent \textit{Proof}.
We observe that the labelling $\mathcal{L}_{I}$ is a Dung's \textit{complete} labelling. All the initial arguments are labelled \texttt{in} and all the other labels are assigned by the \texttt{in-out-fw} algorithm and therefore they all satisfy the constrains of definition 2.5. In the following we exploit the fact that the \textit{grounded} semantics is the complete semantics minimizing the set of \texttt{in}-labelled arguments. \\

\noindent 1) We prove that \textit{in($\mathcal{L}_{I}$)} $\subseteq$ \textit{in($\mathcal{G}_{AF}$)}.
Since $\mathcal{L}_{I}=$ \texttt{in-out-fw}$(I(AF),\mathcal{L}_{und})$ and $I(AF) \subseteq$ \textit{in($\mathcal{G}_{AF}$)}, then for Proposition 4.4 it is  \textit{in($\mathcal{L}_{I}$)} $\subseteq$ \textit{in($\mathcal{G}_{AF}$)}.\\

\noindent 2) We prove that \textit{in($\mathcal{G}_{AF}$)} $\subseteq$ \textit{in($\mathcal{L}_{I}$)}
By contradiction, if \textit{in($\mathcal{G}_{AF}$)} $\not \subset$ \textit{in($\mathcal{L}_{I}$)}, then $\mathcal{G}_{AF}$ is not the  \textit{complete} labelling minimizing the set of \texttt{in}-labelled argument w.r.t. set inclusion and therefore $\mathcal{G}_{AF}$ is not the \textit{grounded} labelling of $AF$, a contradiction. $\square$ \\

We also prove the following theorem, showing that every \textit{weakly complete} extension is a super-set of the \textit{grounded} semantics.
\\

\noindent \textbf{Theorem 4.1} \textit{Let us consider $AF = \langle Ar,\mathcal{R} \rangle$. For each weakly complete labelling $\mathcal{W}$ of $AF$, it holds that $in(\mathcal{G_{AF}}) \subseteq  in(\mathcal{W})$, where $\mathcal{G_{AF}}$ is the \textit{grounded} labelling of $AF$}.\\

\noindent \textit{Proof}.
We notice that every \textit{weakly complete} labelling $\mathcal{W}$ accepts all the initial arguments of the framework, and therefore $I(AF) \subseteq in(\mathcal{W})$. 
Since $\mathcal{G}_{AF}=\mathcal{L}_{I}=$ \texttt{in-out-fw}$(I(AF),\mathcal{L}_{und})$ and $I(AF) \subseteq$ \textit{in($\mathcal{W}$)}, then for Proposition 4.4 it is  \textit{in($\mathcal{G}_{AF}$)} $\subseteq$ \textit{in($\mathcal{W}$)}. $\square$
\\

\noindent The following proposition is the mechanism to compute any \textit{weakly complete} labelling.\\

\noindent \textbf{Proposition 4.6} \textit{Given an argumentation framework $AF = \langle Ar,\mathcal{R} \rangle$ and a \textit{weakly complete} labelling $\mathcal{W}_{AF}$, if argument $g \in Ar$ is labelled} \texttt{undec} \textit{in $\mathcal{W}_{AF}$ and the} \texttt{in-out-fw}$(\{g\},\mathcal{W}_{AF})$ \textit{algorithm ends without generating an inconsistency, than the resulting labelling $\mathcal{W}_{AF}^g$ is a valid \textit{weakly complete} labelling}.\\

\noindent \textit{Proof}.
By Proposition 4.2, the labelling $\mathcal{W}_{AF}^g$ is such that \textit{in$(\mathcal{W_{AF}})$} $\subseteq$ \textit{in$(\mathcal{W}_{AF}^g)$}. The new labels \texttt{in} or \texttt{out} in $\mathcal{W}_{AF}^g$ that were  \texttt{undec} in $\mathcal{W_{AF}}$ are a consequence of the \texttt{in-out-fw}$(\{g\},\mathcal{W_{AF}})$ algorithm. Therefore the new labels are a necessary consequence of accepting the labelling in $\mathcal{W_{AF}}$ (valid by hypothesis) and accepting $g$.
Therefore we just need to prove that the label \texttt{in} assigned to the \textit{ground} argument $g$ in $\mathcal{W}_{AF}^g$ is valid in a \textit{weakly complete} labelling. Since $g$ was labelled \texttt{undec} in $\mathcal{W}_{AF}$, at least one of its attackers was \texttt{undec} and none of them \texttt{in}. The \texttt{in-out-fw}$(\{g\},\mathcal{W_{AF}})$ algorithm might change some or all the undecided attackers of $g$ to \texttt{out}, but not to \texttt{in}, otherwise an inconsistency would be found violating our hypothesis.
Therefore after the application of \texttt{in-out-fw}$(\{g\},\mathcal{W}_{AF})$ all the attackers of $g$ are either labelled \texttt{out} or \texttt{undec} and therefore for the \textit{ub-condition} \textit{3.1b} the label \texttt{in} assigned to $g$ is valid in a \textit{weakly complete} labelling. $\square$ \\

According to propositions 4.6 and 4.2, every consistent application of the \texttt{in-out-fw} algorithm finds a new valid \textit{weakly complete} labelling and the set of \texttt{in}-labelled arguments of each new labelling strictly contains the previous one, since the new application of the \texttt{in-out-fw}$(\{g\})$ does not change the previously assigned labels (Proposition 4.2) and it adds (at least) $g$ to the set of \texttt{in}-labelled arguments. 

Proposition 4.6 also solves the credulous acceptance problem for \textit{weakly complete} semantics. Given an argument $a$, the problem is whether there is at least one valid \textit{weakly complete} labelling where $a$ is accepted. 
In order to check this, the \textit{grounded} labelling $\mathcal{G}_{AF}$ can be computed first. If the argument is accepted or rejected by \textit{grounded} semantics, the credulously acceptance of $a$ is decided, since all \textit{weakly complete} labellings include $\mathcal{G}_{AF}$. For the arguments left undecided in $\mathcal{G}_{AF}$, the credulous acceptance can be checked by computing the \texttt{in-out-fw}($\{a\},\mathcal{G}_{AF}$) and by verifying if the algorithm returns an inconsistency. If not, by Proposition 4.6 we have found a valid \textit{weakly complete} labelling where $a$ is accepted\\

\noindent \textbf{Theorem 4.2}. \textit{Given $AF = \langle Ar,\mathcal{R} \rangle$ and its grounded labelling $\mathcal{G}_{AF}$, let us consider an argument $a$ so that $\mathcal{G}_{AF}(a)=$}\texttt{undec}. \textit{Then $a$ is credulously accepted by \textit{weakly complete} semantics iff} \texttt{in-out-fw}$(\{a\},\mathcal{G}_{AF}$) \textit{does not generate an inconsistency}.\\

\textit{Proof}.

$\leftarrow$ If the \texttt{in-out-fw}$(\{a\},\mathcal{G}_{AF}$) algorithm ends without inconsistency then for Proposition 4.6 there is a valid \textit{weakly complete} labelling containing argument $a$ and therefore $a$ is credulously accepted.\\

$\rightarrow$ By contradiction, we prove that if \texttt{in-out-fw}$(\{a\},\mathcal{G}_{AF}$) generates an inconsistency, then $a$ is not credulously accepted. If \texttt{in-out-fw}$(\{a\},\mathcal{G}_{AF}$) generates an inconsistency, this means that assigning the label \texttt{in} to argument $a$ necessarily generates a contradiction with an argument labelled \texttt{in} by $\mathcal{G}_{AF}$. Therefore in any labelling $\mathcal{L}_{a}$ where $a$ is accepted at least one argument in $in(\mathcal{G}_{AF})$ has to be rejected and therefore \textit{in$(\mathcal{G}_{AF})$} $\not \subseteq$ \textit{in$(\mathcal{L}_{a})$}. However, since for all  \textit{weakly complete} labellings $\mathcal{W}$ it is \textit{in$(\mathcal{G}_{AF})$} $\subseteq$ \textit{in$(\mathcal{W})$}, then $\mathcal{L}_{a}$ is not a \textit{weakly complete} labelling and $a$ is not credulously accepted. $\square$   \\

The following Proposition provides the computational class of the credulous acceptance decision problem of \textit{weakly complete} semantics.\\

\noindent \textbf{Proposition 4.5}. \textit{The computational class of the credulous acceptance problem of \textit{weakly complete} semantics is $PTIME$}.\\

\noindent \textit{Proof}.
If argument $a$ is labelled \texttt{in} or \texttt{out} by the \textit{grounded} semantics, the decision problem is solved in polynomial time since this is the computational class of the \textit{grounded} semantics. If $a$ is left undecided by \textit{grounded} semantics, then it is sufficient to compute \texttt{in-out-fw}$(\{a\},\mathcal{G}_{AF})$ that performs a graph visit of all the nodes reachable from $a$, and this visit has a complexity of $PTIME$ as well. $\square$  \\

We stress the difference with the credulous acceptability problem of \textit{complete} semantics, that is $NP$-\textit{complete}. Under \textit{complete} semantics, the constraint of admissibility requires an accepted argument $a$ to have all its attackers labelled \texttt{out}. An application of the \texttt{in-out-fw}$(\{a\})$ algorithm might not guarantee it, and testing the credulous acceptance in general would require to apply \texttt{in-out-fw} to other arguments that are supposed to \textit{defend} $a$, potentially generating a combinatorial number of applications of the algorithm. 
On the contrary, under \textit{weakly complete} semantics we do not need to check if all the attackers of $a$ are \texttt{out}, since it is sufficient that \texttt{in-out-fw}$(\{a\})$ does not change one of the attackers of $a$ to \texttt{in}.
The credulous acceptance of $a$ is therefore \textit{responsibility} of argument $a$ only, it does not depend on other arguments that might be required to defend $a$.

We now describe our algorithm to compute \textit{weakly complete} semantics, consisting in the repeated application of Proposition 4.6 on different sequences of arguments.
The ground-based algorithm for \textit{weakly complete} labelling is described below. 

\vspace{2mm}
\begin{algorithm}[H]
\SetAlgoLined
\DontPrintSemicolon
\textbf{Input}:{$AF = \langle Ar,\mathcal{R} \rangle$}\;
\textbf{Outputs}: {$W_{l}$ = set of \textit{weakly complete} labellings of $AF$} \;
\;
$L \leftarrow $ \textbf{Grounded}(AF)\;
$W_{l} \leftarrow W_{l} \cup L$ \;
$I \leftarrow \{a \in Ar | \nexists x \in Ar : R(x,a) \in \mathcal{R}\}$\;
$V \leftarrow \emptyset$\;
\textbf{compute}$(W_{l},I,V)$ \;
\;
\textbf{function compute}$(L,Gr,V)$ \;
\Indp
$Candidates \leftarrow  \{x \in Ar | L(x) = undec \land x \not \in V \}$\;
\For{g $\in$ Candidates}
{
    $L^{'} \leftarrow $\textbf{in-out-fw}(\{g\},L) \;
    $V \leftarrow V \cup \{g\}$ \;
    
    \If {$L^{'} \neq$ \textbf{inconsistency}}  {$W_{l} \leftarrow W_{l} \cup L^{'}$ \;
    $Gr \leftarrow  Gr \cup \{g\}$ \;
    $\textbf{compute}(L',Gr,V)$
}}
\caption{The ground-based algorithm}
\end{algorithm}
\vspace{2mm}

The algorithm returns $\mathcal{W}_{l}$, the set of all the \textit{weakly complete} labellings of an argumentation framework. For each labelling $\mathcal{L}$, the variable $Gr$ represents the sequence of arguments used as \textit{ground} in the successive calls of the \texttt{in-out-fw} algorithm used to generate $\mathcal{L}$.\\
First, the algorithm computes \textit{grounded} semantics and add the initial arguments of the framework to the set of grounds $Gr$. Then, it calls a recursive function $\textbf{compute}$. The function tries to apply the \texttt{in-out-fw} algorithm to each undecided argument $g$. If the \texttt{in-out-fw}$(\{g\})$ algorithm does not generate inconsistency, a new labelling is found, $g$ is added to the list of grounds $Gr$ for the new labelling and the algorithm recursively calls itself on the labelling just identified. If an inconsistency is generated, the recursive branch is terminated. 
The algorithm terminates when there are no more undecided arguments to try. The variable $V$ is used to store the list of nodes already used as \textit{ground} in each recursive step.
We describe the functioning of the ground-based algorithm with the following example.\\

\noindent \textbf{Example 4.1}
Let us consider figure 5, that shows the application of the ground-based algorithm to the graph $G_{6}$.
Table 2 shows the five labellings found. Note how the set of initial arguments for the graph $G_{6}$ is empty.

\begin{table}[h]
\centering
\begin{tabular}{l l l}
\hline
\textbf{Labelling} & \textbf{Grounds} & \textbf{Comment}\\
\hline
$L_{1}$ & $\emptyset$ & Grounded Labelling \\
$L_{2}$ & $\{a\}$ &  Complete \\
$L_{3}$ & $\{b\}$ &  Complete\\
$L_{4}$ & $\{d\}$ & Non complete labelling \\
$L_{5}$ & $ \{b,d\}$ & Same as $L_{3}$ \\
\hline
\end{tabular}
\caption{Weakly complete labellings for graph $G_{6}$.}
\end{table}

After finding the \textit{grounded} labelling ($L_{1}$) with all the arguments undecided, the algorithm selects as a ground the argument $a$ and \texttt{in-out-fw}$(\{a\},L_{1})$ algorithm finds the labelling $L_{2}$. All the attempts to extend $L_{2}$ fail, since selecting as ground argument $c$, $d$ or $e$ lead to an inconsistency generated by the cycle of three arguments. This ends this recursive branch of the execution.
The algorithm now selects $b$ as a ground and it finds $L_{3}$, a stable labelling with no undecided arguments that therefore ends this branch of the computation. 
Any attempt to select as ground $c$ or $e$ fails due to the inconsistency of the odd-length cycle.
The selection of $d$ as a ground generated the labelling $L_{4}$, that is \textit{weakly complete} but not \textit{complete}. Attempts to extend $L_{4}$ using $a$ as ground fails, since $b$ becomes \texttt{out}, $c$ \texttt{in} and it conflicts with $d$. Also the choice of $c$ fails for the same conflicts. Choosing $b$ as a ground generates another valid labelling $L_{5}$. This is the same labelling as  $L_{3}$, showing that the ground $b$ was enough to identify this labelling. \\

\begin{figure}[h]
\begin{center}
\includegraphics[scale=0.52]{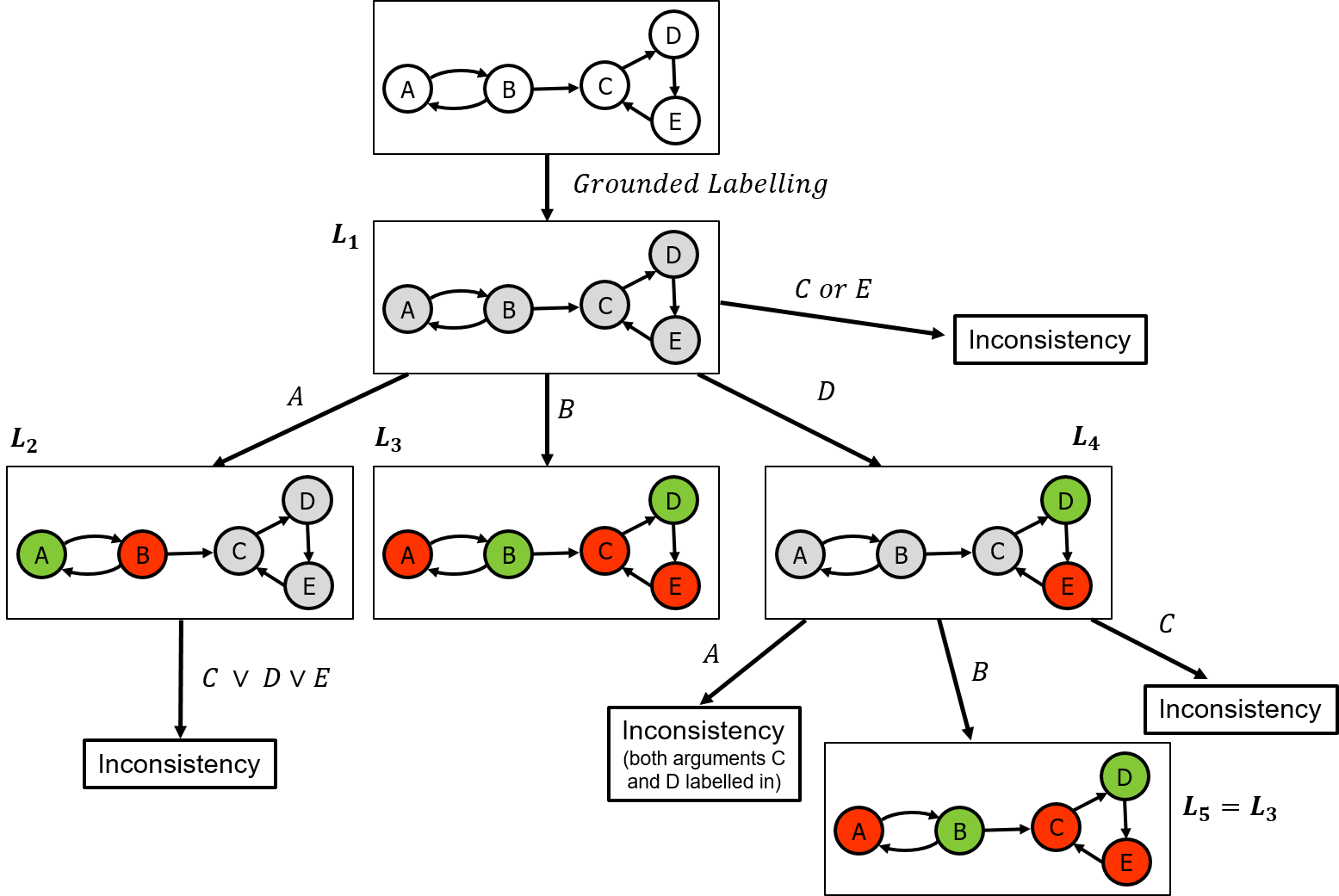}
\caption{Two Argumentation Graphs $G_{1}$ (left) and $G_{2}$ (right)}
\label{fig1}
\end{center}
\end{figure}

\noindent The following theorem proves that the algorithm is correct.\\

\noindent \textbf{Theorem 4.3} \textit{Given $AF = \langle Ar,\mathcal{R} \rangle$, the ground-based algorithm returns all and only the \textit{weakly complete} labellings of $AF$}.\\

\noindent \textit{Proof}. 
$\rightarrow$ Any labelling found by the algorithm is \textit{weakly complete}. This is proved using Proposition 4.6, since each labelling found by the algorithm is generated by consistent applications of the \texttt{in-out-fw} algorithm starting from a valid \textit{weakly complete} labelling, using the \textit{grounded} labelling as the starting labelling. \\

$\leftarrow$ 
Let us consider a valid \textit{weakly complete} labelling $\mathcal{W}_{AF}$. By proposition 4.1 we saw that   $\mathcal{W_{AF}}=$ \texttt{in-out-fw}$(\mathcal{W}_{AF},\mathcal{L}_{und})$. 

We show that the labelling identified by \texttt{in-out-fw}$(\mathcal{W}_{AF},\mathcal{L}_{und})$ is included in the output of the ground-based. Since \texttt{in-out-fw}$(\mathcal{W}_{AF},\mathcal{L}_{und})$ does not generate an inconsistency since $\mathcal{W}_{AF}$ is a valid \textit{weakly complete} labelling, we can also compute it by applying the \texttt{in-out-fw} sequentially starting from $\mathcal{L}_{und}$. We first use as grounds all the initial arguments, obtaining the \textit{grounded} labelling, and then all the arguments that are in $in(\mathcal{W_{AF}}) \setminus in(\mathcal{G_{AF}})$ obtaining $\mathcal{W_{AF}}$. 
Note how some of the arguments in $in(\mathcal{W_{AF}}) \setminus in(\mathcal{G_{AF}})$ might not be used for the \texttt{in-out-fw} algorithm since they could have been labelled \texttt{in} by the effect of other \textit{ground} arguments used before them. In this case they are skipped since their constrains have been already generated by previous arguments. Note how these arguments cannot be labelled \texttt{out} by previous application of \texttt{in-out-fw}, otherwise $\mathcal{W}_{AF} \not$ = \texttt{in-out-fw}$(in(\mathcal{W}_{AF}),\mathcal{L}_{und})$, a contradiction. $\square$

\section{Families of weakly complete semantics}\label{sec:famWeaklySemantics}
In this section several families of \textit{weakly complete} semantics are defined. The first three semantics are defined in a similar way in which families of \textit{complete} semantics are defined, i.e. by identifying subset of semantics whose set of \texttt{in}- or \texttt{undec}-labelled arguments satisfy maximality or minimality conditions.
The last two semantics are called \textit{undecidedness blocking} semantics.

We have seen how several \textit{weakly complete} labellings can be generated by deciding when and how often to use the \textit{ub-condition} \textit{3.1b} instead of the \textit{undecidedness propagating} (\textit{up-condition}) \textit{3.1c}. In the \textit{undecidedness blocking} semantics the use of the \textit{ub-condition} \textit{3.1b} is constrained by the postulates of the semantics. These semantics mimic the behaviour of \textit{ambiguity blocking} semantics of defeasible logic, where ambiguity is always blocked and never propagated.
The first semantics, called \textit{undecidedness blocking preferred} semantics, is the semantics where the \textit{ub-condition} is used \textit{as much as possible} and \textit{as earlier as possible}. The second, called \textit{undecidedness blocking grounded} semantics, is a single-status  non-admissible variant of \textit{grounded} semantics retaining the majority of properties of its Dung's counterpart.

\subsection{Weakly preferred semantics} 
\noindent\textbf{Definition 5.1} 
\textit{Given an argumentation framework $AF = \langle Ar,\mathcal{R} \rangle$, the \textit{weakly preferred labelling} of $AF$ is the \textit{weakly complete} labelling where the set of} \texttt{in}\textit{-labelled arguments is maximal w.r.t. set inclusion}.\\

\noindent We have seen how the set of \texttt{in}-labelled arguments increases strictly monotonically each time the \texttt{in-out-fw} algorithm is applied without inconsistency. Since each application of the \texttt{in-out-fw}$(\{a\})$ corresponds to an application of the \textit{ub-condition} \textit{3.1b} on argument $a$, a \textit{weakly preferred} labelling is the labelling where the \textit{ub-condition} is used as \textit{much as possible}: no more undecided arguments can be promoted to the label \texttt{in} without causing an inconsistent application of the \texttt{in-out-fw} algorithm. 
In terms of the ground-based algorithm, the \textit{weakly preferred} labellings are the ones identified by the terminal nodes of each recursive branch.
In the graph $G_{6}$ of figure 5, the \textit{weakly preferred} labellings are $\mathcal{L}_{2}$ and $\mathcal{L}_{3}=\mathcal{L}_{5}$, that are the terminal nodes of the recursion.

\subsection{Weakly Grounded semantics}
\noindent \textbf{Definition 5.2} \textit{
Given an argumentation framework $AF = \langle Ar,\mathcal{R} \rangle$, the \textit{weakly grounded} labelling of $AF$ is the \textit{weakly complete} labelling where the set of} \texttt{undec}\textit{-labelled arguments is maximal w.r.t. set inclusion.}\\

\noindent Since every \textit{weakly complete} labelling $\mathcal{W}$ is found by applying the\texttt{in-out-fw} algorithm to Dung's \textit{grounded} labelling and, and since at each subsequent consistent application of the \texttt{in-out-fw} algorithm the set $in(\mathcal{W})$ is stricly increased and therefore $undec(\mathcal{W})$ is reduced, the \textit{weakly grounded} semantics is Dung' \textit{grounded} semantics.
In terms of the ground-based algorithm, the \textit{grounded} semantics is obtained by using as \textit{ground} arguments the initial arguments of the framework (Proposition 4.4).

\subsection{Weakly Stable Semantics}
\noindent \textbf{Definition 5.3} \textit{
Given an argumentation framework $AF = \langle Ar,\mathcal{R} \rangle$, the \textit{weakly stable} labelling of $AF$ is the \textit{weakly complete} labelling where the set of} \texttt{undec}\textit{-labelled arguments is empty}.\\

\noindent The \textit{weakly stable} labelling coincides with Dung's \textit{stable} labelling. Note how a stable labelling is always a \textit{weakly preferred} labelling: since no undecidedness is present in the labelling the \textit{ub-condition} cannot be used further.

\subsection{The undecidedness-blocking semantics}
\textit{Undecidedness blocking} semantics (\textit{ub}-semantics) are \textit{weakly complete} semantics where the \textit{undecidedness blocking} condition is used \textit{as much as possible} and \textit{as earlier as possible}. 
As it happens in the \textit{weakly preferred} semantics, saying that the \textit{ub-condition} is used \textit{as much as possible} means that it cannot be used further. However, this does not guarantee that the undecidedness is blocked \textit{as earlier as possible}. Figure 6 shows two \textit{weakly complete} labelling for a graph (a third valid labelling exists, and it is the \textit{grounded} labelling). The source of undecidedness is the self-attacking argument $a$. Indeed, in the labelling on the left undecidedness is blocked at argument $b$, \textit{earlier} than in the labelling on the right, where it is blocked at argument $c$. However, both of them are \textit{weakly preferred} labelling, since both of them maximise the set of \texttt{in}-labelled arguments.

\begin{figure}[h]
\begin{center}
\includegraphics[scale=0.55]{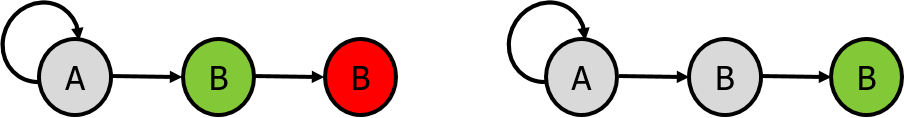}
\caption{Two \textit{weakly complete} labellings. The one on the left is also a ub-grounded labelling}
\label{fig1}
\end{center}
\end{figure}

The concept of \textit{as earlier as possible} has to consider the topological order of the arguments of the argumentation framework.
In \textit{ub-semantics}, a sufficient condition to block undecidedness is the following: if argument $a$ has an attacker $b$, and there is no directed path back from $a$ to $b$, then in all the labellings where $b$ is undecided the \textit{undecidedness blocking} condition \textit{3.1b} is always applied to $a$ and the attack from $b$ is \textit{de facto} neglected. In absence of other \texttt{in}-labelled attackers, $a$ is therefore labelled \texttt{in}.
This leaves open the problem of how to block undecidedness \textit{as earlier as possible} inside a cyclic strongly connected component where no topological order is defined. Different strategies lead to different variants of \textit{ub-semantics}.

The importance of the topological order makes convenient to model \textit{ub-semantics} using an instance of the SCC-recursive schema \cite{baroni2005scc}, that we quickly describe. 
We consider the graph $G_{SCC}$ composed by the strongly connected components of the original graph $G$. $G_{SCC}$ is a direct acyclic graph for which it is possible to define a topological order. 
Using the SCC-recursive schema, the extensions of a semantics are computed following the topological order of the $G_{SCC}$ graph. Initial strongly connected components are labelled using a base function $\mathcal{B}$. A non-initial strongly connected component $S$ is labelled considering external attacks from argument belonging to SCCs preceding $S$ in the topological order of the graph $G_{SCC}$ and therefore already labelled by the semantics.
In the general SCC-recursive scheme, one has to consider attacks from arguments in the extension (labelled \texttt{in}) and provisionally defeated arguments (labelled \texttt{undec}). 
In \textit{ub-semantics}, in order to implement the \textit{undecidedness blocking} constrain described above, when labelling a non-initial strongly connected component $S$ we neglect attacks to arguments in $S$ from undecided arguments external to $S$, and only attacks from \texttt{in}-labelled arguments are considered.  Undecidedness results blocked rather than propagated into $S$.
The \textit{ub-labelling} function corresponding to \textit{ub-semantics} can be formalized as follows.\\

\noindent \textbf{Definition 5.4} \textit{Let us consider the argumentation framework $AF=\langle Ar,\mathcal{R} \rangle$, and a labelling function $\mathcal{L}$. The \textit{ub}-labelling is identified by the function $\mathcal{L}_{ub}$ defined as follows:}
\begin{itemize}

\item if $|SCCS_{AF}|=1$, $\mathcal{L}_{ub} = \mathcal{L}$ 
\item otherwise, $\forall S \in SCCS_{AF}$, it is
$$
\mathcal{L}_{ub}=
\begin{cases}
\mathcal{L}_{ub}, \;\;\;  \forall x \in S \setminus S^{+}_{in} \\
\texttt{out}, \;\;\; \forall x \in S^{+}_{in}
\end{cases}
$$

\end{itemize}

\noindent \textit{where $SCCS_{AF}$ is the set of all the strongly connected components of $AF$ and $S^{+}_{in}$ is the set of arguments in $S$ externally attacked by an} \texttt{in}\textit{-labelled argument:
$S^{+}_{in}=\{a\in S \; | \; \exists b \not \in S : R(b,a) \land \mathcal{L}_{ub}(b)=$}\texttt{in}$\}$.\\

\noindent We require the base function $\mathcal{L}$ to return labellings that are \textit{weakly complete}. Indeed, if $\mathcal{L}$ returns \textit{weakly complete} labellings then $\mathcal{L}_{ub}$ returns \textit{weakly complete} labellings as well. This follows from the fact that the $\mathcal{L}_{ub}$ labelling is generated by applying the \textit{weakly complete} labelling function $\mathcal{L}$ on strongly connected components (or restrictions of such components) of $AF$, while the \textit{undecidedness blocking} mechanism that neglects attacks from undecided arguments to a non-initial strongly connetected component is justified by the \textit{ub-condition} \textit{3.1b} of \textit{weakly complete} labellings. 

The following Proposition shows how the undecidedness is not propagated outside the strongly connected component where it was generated, and acyclic arguments are labelled either \texttt{in} or \texttt{out}.\\

\noindent \textbf{Proposition 5.1}. \textit{Let us consider the argumentation framework $AF=\langle Ar,\mathcal{R} \rangle$. If an argument $a \in Ar$ is acyclic, then $\mathcal{L}_{ub}(a) \neq$} \texttt{undec}.\\

\noindent \textit{Proof}. 
Since $a$ is not part of a cycle, it is a strongly connected component of the original graph. Therefore, according to the definition of \textit{ub-semantics}, the labelling of $a$ is either the application of the base function on $a$ alone - and therefore $a$ is labelled \texttt{in} since the base function is \textit{weakly complete} - or it is labelled \texttt{out} if the argument has one attacker labelled \texttt{in}. $\square$

\subsubsection{Undecidedness Blocking Grounded Semantics}

\noindent \textbf{Definition 5.5}. The \textit{undecidedness-blocking grounded semantics $\mathcal{G}_{ub}$ is the \textit{ub-semantics} using as base function the \textit{grounded} semantics $\mathcal{G}$}.\\

\noindent The use of the \textit{grounded} semantics as base function implies that the arguments part of a cyclic strongly connected component $S$ with no external attacks from \texttt{in}-labelled arguments (i.e. $S^{+}_{in}=\emptyset$) are all labelled \texttt{undec}, as it happens with \textit{grounded} semantics.
However, the use of the SCC-recursive scheme of \textit{ub-semantics} guarantees that the acyclic arguments are labelled \texttt{in} or \texttt{out} by \textit{ub-grounded} semantics (Proposition 5.1), and therefore the undecided label is not propagated (but blocked) outside the cycle where it was generated. Unattacked conflicting arguments are unresolved conflicts generating undecidedness, but this undecidedness is not propagated to other arguments not part of the conflict.
Undecidedness is therefore blocked \textit{as earlier as possible} (in the topological sense) outside cycles.
An example of \textit{ub-grounded} semantics is shown in figure 6. Only the \textit{weakly complete} labelling on the left is a \textit{ub-grounded} labelling. The source of undecidedness is the self-attacking argument $a$. Indeed, in the labelling on the left undecidedness is blocked \textit{earlier} than in the labelling on the right. 
\\

\noindent \textbf{Proposition 5.2} \textit{The ub-grounded labelling always exists and it is unique}.
\vspace{3mm}

\noindent \textit{Proof}. Building the \textit{ub-grounded} labelling equates to a series of application of the \textit{grounded} semantics on strongly connected components or restriction of strongly connected components of the argumentation framework.
Therefore the thesis is proven by relying on the fact that \textit{grounded} semantics always exists and it is unique $\square$.

\subsection{Undecidedness Blocking Preferred Semantics}
The second \textit{ub-semantics} aims to use the \textit{ub-condition} not only \textit{as earlier as possible} but also \textit{as much as possible}.
In the \textit{ub-grounded} semantics, undecidedness is only blocked for arguments outside cyclic strongly connected components, but it is not blocked inside the components. 
However, a single strongly connected component can have multiple \textit{weakly complete} labellings in which undecidedness is blocked, as shown in figure 7.

While the notion of \textit{as much as possible} has been already captured by \textit{weakly preferred} labellings, it is more challenging to decide when a labelling uses the \textit{ub-condition} \textit{earlier} than another. Referring to figure 7, which one is \textit{earlier}? 
In the definition of \textit{ub-semantics} the concept of \textit{earlier} has been mapped to the notion of topological order, that is nevertheless not applicable inside a single strongly connected component where no topological order defined.

We might decide that there is no point in considering an ordering among elements of the same strongly connected component and use the \textit{weakly preferred} labelling as base function, since it guarantees that the \textit{ub-condition} is used \textit{as much as possible}. This solution is not completely satisfactory. Referring to figure 7, the two labellings are both \textit{weakly preferred} labellings, but intuitively undecidedness is blocked earlier in the graph on the right (where undecidedness is blocked at argument $b$) than in the graph on the left (where undecidedness is blocked at $c$). Note how $b$ is also \textit{closer} to $a$, the source of the unresolved conflict generating the undecided label.

How can we capture the above intuition?
Since the effect of the \textit{ub-condition} is to assign  the label \texttt{in} to otherwise undecided arguments, we might  hypothesize that it is a question of selecting the labelling minimizing the set of undecided labels. However, this does not capture the property we want, since both the labellings of figure 7 minimize the set of undecided arguments.
What we need to define is an ordering among arguments of a strongly connected component based not on the topology of the graph but rather on the relation among their labels. The ordering is formalized as follows: \\ 

\begin{figure}[h]
\begin{center}
\includegraphics[scale=0.55]{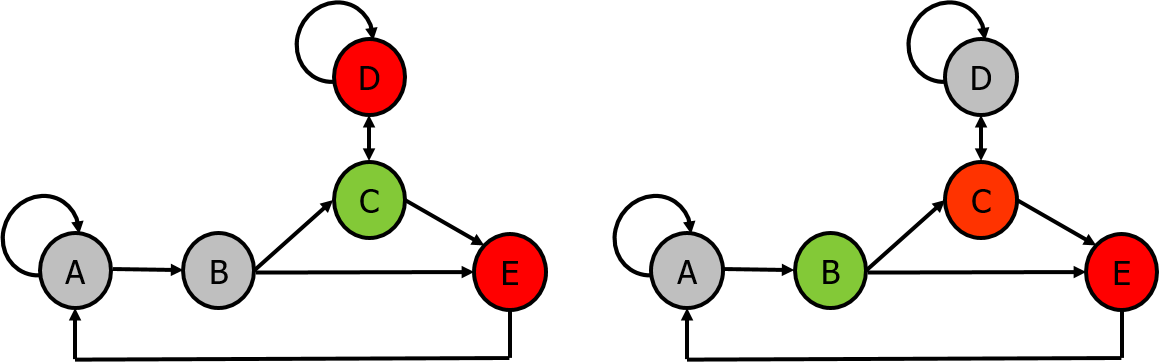}
\caption{Two \textit{weakly complete} labellings. The undecidedness blocking condition is applied to argument $c$ on the left and $b$ on the right.}
\label{fig1}
\end{center}
\end{figure}

\noindent \textbf{Definition 5.6 (semantic precedence)}. \textit{ Given $AF=\langle Ar,\mathcal{R} \rangle$, we define $\mathcal{P}\subset Ar \times Ar$, a binary relation over arguments so that $(a,b)\in \mathcal{P}$ iff $a$ and $b$ are part of the same strongly connected component, $a$ is at least credulously accepted by \textit{weakly complete} semantics and for all the \textit{weakly complete} labellings $\mathcal{L}$ of $AF$ it holds that $(\mathcal{L}(a)=$}\texttt{in}) $ \rightarrow (\mathcal{L}(b)\neq$ \texttt{undec}). \textit{We define \textit{\textbf{semantic precedence}} the binary relation $\succ_{s}\subset Ar \times Ar$ so that $a \succ_{s} b$ iff $(a,b) \in \mathcal{P} \land (b,a) \not\in \mathcal{P}$}.\\

The \textit{semantic precedence} relation $\succ_{s}$ captures the idea that the acceptance of an argument affects the status of another, but not vice versa.
Referring to the graph in figure 7, it is $b \succ_{s} c$, $c \succ_{s} d$ and $c \succ_{s} e$.
The idea is that in order to apply the \textit{ub-condition as earlier as possible}, we should consider the relation $\succ_{s}$. The \textit{ub-condition} should be applied to argument $a$ before $b$ if $a \succ_{s} b$. 

In order to formalize this idea, we consider the set $I_{p}$ of all the arguments for which there is no argument preceding them, that is
$I_{p}=\{a\in Ar| \nexists b \in Ar $ so that $ b \succ_{s} a\}$.
We now consider the \textit{weakly preferred} labelling $\mathcal{L}$ that maximises the set of arguments accepted in $I_{p}$. This labelling uses the \textit{ub-condition} as much as possible (since it is a \textit{weakly preferred} labelling) but it also considers the semantic precedence and it first accepts arguments for which there are no arguments semantically preceding them. 
A \textit{ub-preferred} semantics is the \textit{ub-semantics} using as base function $\mathcal{L}$. Formally:\\

\textbf{Definition 5.7} \textit{Let us consider an argumentation framework $AF=\langle Ar,\mathcal{R} \rangle$ and the \textit{weakly preferred} labelling function $\mathcal{W}_{pr}$. We define the labelling function $\mathcal{L}$ as the function returning all the labellings for which \textit{in}($\mathcal{W}_{pr}$) $\cap$ $I_{p}$ is maximal w.r.t. set inclusion. The \textit{ub-preferred} semantics $\mathcal{L}_{ub_{pr}}$ is the \textit{ub-semantics} where the base function is $\mathcal{L}$.}\\

Referring to figure 7, only the labelling on the right is a \textit{preferred ub-labelling}, since $I_{p}=\{b\}$. Referring to figure 6, only the graph on the left is a preferred ub-labellings, since $I_{p}=\{b\}$ as well. In graph $G_{5}$ of figure 3, $I_{p}=\{a,b\}$ and therefore only the two Dung's \textit{complete preferred} labellings are \textit{ub-preferred} labellings. 


\section{Discussion and Properties}
In this section we provide a principle-based analysis of \textit{weakly complete} semantics w.r.t. a set of properties commonly referred in literature and collected from \cite{baroni2011introduction} and \cite{van2017principle}.
Table 3 summarizes the results. The first four columns refers to Dung's \textit{complete, grounded, preferred} and \textit{stable} semantics, and they serve as a comparison. We remind how \textit{weakly stable} and \textit{weakly grounded} semantics coincide with Dung's \textit{stable} and \textit{grounded} semantics. The following four columns refer respectively to \textit{weakly complete}, \textit{weakly preferred},  \textit{undecidedness blocking grounded} and \textit{undecidedness blocking preferred} semantics. The last three columns refer to the \textit{weakly admissible} version of \textit{complete, grounded} and \textit{preferred} semantics proposed by Baumann et al. \cite{bbu1}. For such semantics we report the results of the principle-based analysis conducted in \cite{bub_principle}.

\begin{table}[]
\centering
\footnotesize
\caption {Properties of \textit{weakly complete},  \textit{ub-semantics}}
\hskip-1.0cm
\begin{tabular}{|c|c|c|c|c|c|c|c|c|c|c|c|}
\hline
\textbf{Property}       & \textit{co} & \textit{gr} & \textit{pr} & \textit{st} & \textit{wc} & \textit{$w_{pr}$} & \textit{$ub_{gr}$} & \textit{$ub_{pr}$} & \textit{$co_{bbu}$} & \textit{$gr_{bbu}$} & \textit{$pr_{bbu}$} \\
\hline
Conflict-free  & Yes               & Yes     & Yes  & Yes &Yes &Yes &Yes & Yes & Yes &Yes &Yes  \\\hline
Naiveity  & No               & No     & No  & No &No &No &No & No & 
No&No &No  \\\hline
Admissible     & Yes                & Yes      & Yes   & Yes & No & No & No & No & No & No & No  \\\hline
Reinstatement & Yes               & Yes     & Yes  & Yes &Yes &Yes &Yes & Yes&Yes &Yes & Yes   \\\hline
Rejection      & Yes               & Yes     & Yes  & Yes &Yes &Yes &Yes &Yes&Yes &Yes & Yes    \\\hline
Directionality & Yes               & Yes     & Yes  & No & Yes & Yes & Yes &    Yes & No& No& ?\\\hline
Abstention     & Yes               & Yes      & No  & No & Yes& Yes& No&No& No& No&No    \\\hline
Cardinality    & $\geq 1$                & 1      & $\geq 1$   &$\geq 0$ & $\geq 1$& $\geq 1$& 1& $\geq 1$& $\geq 1$& $\geq 1$& $\geq 1$   \\\hline
I-maximality   & No                & Yes      & Yes & Yes & No& Yes& Yes&Yes&No&Yes&Yes
\\\hline
Cycle-Homo.   & No                & Yes      & No & No & No& No& Yes&No&No&No&No
\\\hline
\end{tabular}
\end{table}
\subsection{Principle-based Analysis}
All \textit{weakly complete} semantics are \textit{non-admissible}, since they can accept arguments that are not defended from the attacks of \texttt{undec}-labelled arguments. These semantics could be seen as employing a different form of admissibility since they still require an argument to be defended from the attacks of \texttt{in}-labelled arguments but, under some conditions, not from the attacks of \texttt{undec}-labelled arguments. 

\textit{Weakly complete} semantics are \textit{conflict-free}, since if an argument $a$ is accepted, the arguments attacked by $a$ are not (\textit{rejection} condition \textit{3.1b}). Therefore \textit{weakly complete} semantics also satisfy \textit{rejection}, since arguments attacked by at least one \texttt{in}-labelled argument are always labelled \texttt{out} and explicitly rejected.

All \textit{weakly complete} semantics satisfy the \textit{reinstatement} property since, if an argument has all its attackers labelled \texttt{out}, it is necessarily labelled \texttt{in}. However, an additional form of reinstatement is possible in \textit{weakly complete} semantics that is not possible in Dung's \textit{complete} semantics.
Indeed, in \textit{ub-grounded} semantics an argument $a$ defeated by $b$ is fully reinstated even by an argument $c$ rebutting $b$, since the attack of $c$ creates a cycle with $b$ and therefore there is a \textit{weakly complete} semantics where $c$ and $b$ are labelled undecided and the undecided label is blocked using the ub-condition \textit{3.1d} and $a$ is accepted.  

The property of \textit{directionality} is defined as follows (from \cite{baroni2007principle}):

\vspace{3mm}
\noindent \textbf{Definition 6.1}.  \textit{Let us consider a framework $AF=\langle Ar,\mathcal{R} \rangle$ and $\mathscr{L}_{\sigma}(AF)$, the set of all the labelligs of $AF$ according to semantics $\sigma$. We consider $US_{AF}$, the set of  initial strongly connected components of $AF$. The semantics $\sigma$ satisfies directionality iff $\forall U \in US_{AF}, in(\mathscr{L}_{\sigma}(AF_{\downarrow U}))=  in(\mathscr{L}_{\sigma}(AF_{\cap U}))$, where $in(\mathscr{L}_{\sigma}(AF_{\downarrow U}))=\{in(\mathcal{L}) | \mathcal{L} \in \mathscr{L}_{\sigma}(AF_{\downarrow U}) \} $
and $in(\mathscr{L}_{\sigma}(AF_{\cap U}))=\{in(\mathcal{L}) \cap U | \mathcal{L} \in \mathscr{L}_{\sigma}(AF) \}$
}
\vspace{3mm}

The idea is that the justification state of an argument $a$ is affected only by the justification state of the defeaters of $a$ (which in turn are affected by their defeaters and so on).
\vspace{3mm}

\noindent \textbf{Proposition 6.1}. \textit{Weakly complete semantics satisfies directionality}.
\vspace{3mm}

\noindent \textit{Proof}. 
We need to prove that $\forall U \in US_{AF}, in(\mathscr{L}_{\sigma}(AF_{\downarrow U}))=  in(\mathscr{L}_{\sigma}(AF_{\cap U}))$. \\

\noindent 1) $in(\mathscr{L}_{\sigma}(AF_{\downarrow U})) \subseteq  in(\mathscr{L}_{\sigma}(AF_{\cap U}))$. 

\noindent We prove that $\forall \mathcal{L}_{\downarrow U} \in \mathscr{L}_{\sigma}(AF_{\downarrow U}), \; \exists \mathcal{L}_{\cap U} \in \mathscr{L}_{\sigma}(AF_{\cap U})$ so that $in(\mathcal{L}_{\downarrow U}) \subseteq  in(\mathcal{L}_{\cap U})$.
Each labelling $\mathcal{L}_{\downarrow U} \in \mathscr{L}_{\sigma}(AF_{\downarrow U})$ is generated by applying the \texttt{in-out-fw} algorithm to a set of ground arguments $G=\{g_1,..,g_n\} \subseteq U$ (Theorem 4.3).
We consider the labelling  $\mathcal{L} \in \mathscr{L}_{\sigma}(AF)$ generated using as grounds the initial arguments of the framework and the arguments in $G$. 
We observe that none of the applications of the \texttt{in-out-fw} applied to the sequence of arguments $G$ produces an inconsistency. Indeed, according to Proposition 4.2, \texttt{in-out-fw}$(\{g\})$ ($g \in G$) can generate an inconsistency only with another ground argument or itself. The constraints generated by \texttt{in-out-fw}$(\{g\})$ cannot conflict with any other ground argument in $G$ (since $\mathcal{L}_{\downarrow U}$ is a valid \textit{weakly admissible} labelling), and neither they can conflict with a ground arguments external to $U$, since $\mathscr{L}_{\sigma}(AF)$ does not have any other non-initial ground arguments outside $U$. Therefore $in(\mathcal{L}_{\downarrow U}) \subseteq in(\mathcal{L})$ and therefore $in(\mathcal{L}_{\downarrow U}) \subseteq in(\mathcal{L}) \cap U = in(\mathcal{L}_{\cap U})$.
\vspace{2mm}

\noindent 2) $in(\mathscr{L}_{\sigma}(AF_{\downarrow U})) \supseteq  in(\mathscr{L}_{\sigma}(AF_{\cap U}))$.

\noindent We prove that $\forall \mathcal{L}_{\cap U} \in \mathscr{L}_{\sigma}(AF_{\cap U}), \; \exists \mathcal{L}_{\downarrow U} \in \mathscr{L}_{\sigma}(AF_{\downarrow U})$ so that $in(\mathcal{L}_{\downarrow U}) \supseteq  in(\mathcal{L}_{\cap U})$.
This is proved by observing that the constraints of the \texttt{in-out-fw}($\{g\}$) with $g \not \in U$ do not affect the labels of arguments in $U$, since $U$ is initial in $AF$. Therefore, for each labelling $\mathcal{L} \in \mathscr{L}_{\sigma}(AF)$ the set  $in(\mathcal{L}) \cap U$ is included in the set of \texttt{in}-labelled arguments of at least one labelling $\mathcal{L}_{\downarrow U} \in \mathscr{L}_{\sigma}(AF_{\downarrow U})$ and therefore $in(\mathscr{L}_{\sigma}(AF_{\downarrow U})) \supseteq  in(\mathscr{L}_{\sigma}(AF_{\cap U}))$.  $\square$
\vspace{3mm}

All the \textit{ub-semanitcs} satisfies \textit{directionality},  since, as proven in Proposition 55 of \cite{baroni2007principle}, they have been defined using an SCC-recursive schema and they have at least one valid labelling. 
The \textit{weakly preferred} labelling does not satisfy \textit{directionality}. As a counter-example, we consider the floating assignment graph $G_5$ of figure 3 and the initial strongly connected component $\{a,c\}$. If $\mathscr{L}(AF)$ is the set of the three \textit{weakly preferred} labellings of $AF$, it is  $in(\mathscr{L}(AF)) \cap \{a,c\}=\{\{a\},\{c\},\emptyset\}$ and $in(\mathscr{L}(AF_{\downarrow \{a,c\}}))=\{\{a\},\{c\}\}$.

The property \textit{abstention} states that, if an argument $a$ is labelled \texttt{out} in at least one valid labelling and labelled \texttt{in} in at least another, then there must be a valid labelling where $a$ is labelled \texttt{undec}. Among Dung' semantics, only \textit{complete} semantics satisfies it. It can be proved that 
\textit{weakly complete} semantics satisfy \textit{abstention}.

\vspace{3mm}
\noindent \textbf{Proposition 6.2}. \textit{Weakly complete semantics satisfies abstention}.

\vspace{3mm}
\noindent \textit{Proof}. Let us consider an argumentation framework $AF=\langle Ar,\mathcal{R} \rangle$ and $a \in Ar$. If an argument $a$ is \texttt{in} in one \textit{weakly complete} labelling and \texttt{out} in another one, $a$ is necessarily labelled \texttt{undec} by the \textit{grounded} labelling $\mathcal{G}_{AF}$. This is because the set of \texttt{in}-labelled arguments of any \textit{weakly complete} labelling contains $in(\mathcal{G}_{AF})$ and, since \textit{grounded} semantics is unique, all the arguments labelled \texttt{out} or \texttt{in} by \textit{grounded} semantics retain their labels in all the \textit{weakly complete} labellings. Therefore if there is a \textit{weakly complete} labelling where $a$ is labelled \texttt{in} and another where $a$ is labelled \texttt{out}, then there is at least one third valid labelling, the \textit{grounded} labelling, where $a$ is labelled \texttt{undec} and the property is proven. $\square$
\vspace{3mm}

\textit{Weakly preferred} labelling does not satisfy \textit{abstention}, a counter-example is a graph with two rebutting arguments. The same counter-argument is valid for \textit{ub-preferred} semantics, while the single-status \textit{ub-grounded} semantics satisfies it, as Dung's \textit{grounded} semantics does.

Regarding the \textit{cardinality} of each semantics, multiple \textit{weakly complete} labellings of the same graph could exist and at least one (the \textit{grounded} labelling) always exists. The same is for \textit{weakly preferred} semantics and \textit{ ub-preferred} semantics. \textit{Ub-grounded} semantics has exactly one labelling (Proposition 5.2).

A semantics satisfies \textit{I-maximality} if no extension is a strict subset of another. It is indeed satisfied by single status semantics like the \textit{ub-grounded} semantics.
It is not satisfied by \textit{weakly complete} semantics, as it is evident from the ground-based algorithm. Indeed, given a valid \textit{weakly complete} labelling, it could be possible to extend the set of \texttt{in}-labelled arguments by a new consistent application of the \texttt{in-out-fw} algorithm. 

The maximality condition of the set of \texttt{in}-labelled arguments of \textit{weakly preferred} labellings implies that the semantics satisfies  \textit{I-maximality}. Indeed, if the \texttt{in}-set of a labelling $\mathcal{L}_{1}$ includes the \texttt{in}-set of $\mathcal{L}_{2}$, than $\mathcal{L}_{2}$ is not a \textit{weakly preferred} labelling.
The same reasoning applies to  \textit{ub-preferred} semantics since by definition \textit{ub-preferred} labellings are all \textit{weakly preferred} labellings. 

The property of \textit{cycle-homogeneity} is introduced in this paper. A semantics satisfies this property iff all the arguments in an argumentation framework composed by a single cycle are labelled in the same way in all the valid labellings of the semantics. Formally:
 
\vspace{3mm}
\noindent \textbf{Definition 6.2}. \textit{Let us consider an argumentation framework $AF=\langle Ar,\mathcal{R} \rangle$ composed by a single cycle of arguments, a semantics $\sigma$ and $\mathscr{L}_{\sigma}(AF)$ the set of labellings of $AF$ generated by $\sigma$. The semantics $\sigma$ satisfies the principle of cycle-homogeneity iff $\forall a \in Ar$, $\nexists \mathcal{L}_1,\mathcal{L}_2 \in \mathscr{L}_{\sigma}(AF)$ so that $\mathcal{L}_1(a) \neq \mathcal{L}_2(a)$}.
\vspace{3mm}

It is a well-known behaviour of multi-status Dung' semantics to label unattacked cycles of arguments in a different way depending on the length of the cycle. An odd-length cycle has an empty \textit{complete} extension, while an even-length cycle has multiple valid labellings. This peculiar way of assigning the acceptability status to odd-length cycles has been indicated as “puzzling” by Pollock \cite{pollock2001defeasible}. Among Dung' semantics, the property of \textit{cycle-homogeneity} is satisfied only by \textit{grounded} semantics, where all the arguments are deemed undecided. This behaviour is retained in the \textit{ub-grounded} semantics.

\begin{table}[]
\begin{tabular}{|c|p{10cm}|}
\hline
\textbf{Semantics} & \textbf{Undecidedness Blocking Constraints}                            \\\hline      
Grounded  w.a.         & 
- No undecidedness blocking.
\\\hline
Preferred w.a.       & 
- Undecidedness blocked \textit{as much as possible}.

- \textit{weakly complete labellings} with maximal \texttt{in}-set

- Terminal nodes of the ground-based recursion tree
\\\hline 
UB-Grounded       & - Undecidedness blocked \textit{as earlier as possible} only outside cyclic strongly connected components  
\\\hline
UB-Preferred      & - Undecidedness blocked \textit{as earlier as possible} and \textit{as much as possible}      \\\hline                              
\end{tabular}
\caption{Conditions for \textit{weakly complete} semantics}
\end{table}

\subsection{Generating Argumentation Semantics using Undecidedness Blocking }
The ground-based algorithm presented in section 4 shows how any \textit{weakly complete} labelling (therefore including any \textit{complete} labelling) can be generated by repeatedly applying the \texttt{in-out-fw} algorithm.
Labellings are build in a sequential way by multiple applications of the \texttt{in-out-fw} algorithm, and each application of the algorithm is a tentative to reduce the set of undecided arguments by blocking undecidedness at the argument used as ground for the \texttt{in-out-fw} algorithm.

The ground-based algorithm unifies the computation of all \textit{weakly complete} semantics, therefore including Dung’s \textit{complete} semantics. As such, it provides a way of interpreting the different semantics as a gradual effort to block undecidedness.

On the one end of the spectrum there is the \textit{grounded} semantics, where undecidedness is not blocked, and on the other side \textit{stable} semantics, where undecidedness is not present.

Every \textit{weakly complete} labelling is a superset of the \textit{grounded} semantics, that represents a necessary baseline for each acceptability strategy. 
\textit{Grounded} semantics accepts only initial arguments and all the arguments defended directly or indirectly by initial arguments. The \texttt{in-out-fw} algorithm is applied only to all the initial arguments of the framework, accepted to satisfy the \textit{admissibility} condiiton \textit{3.1a}. The \textit{undecidedness blocking} condition \textit{3.1d} is not applied to any non-initial undecided arguments, making the \textit{grounded} semantics is the only truly \textit{undecidedness propagating} semantics, since the \textit{ub-condition} is not used in generating the labelling.

All the other labellings are obtained by repeatedly apply the \textit{ub-condition} on some undecided arguments and propagating the necessary constraints. They are therefore all employing some \textit{undecidedness blocking} mechanisms, where the set of undecided arguments is somehow reduced by tentatively accepting some of them. 
However, (1) the arguments on which the \textit{ub-condition} can be used, (2) when and how often  it is used and (3) the conditions to accept as valid the resulting labelling define the different \textit{weakly complete} semantics. Table 4 show the \textit{undecidedness blocking} strategies of the families of \textit{weakly complete} semantics presented in this paper.

In \textit{weakly complete} semantics any argument $a$ can be selected as ground (thus blocking undecidedness at $a$) as long as \texttt{in-out-fw}$(\{a\})$ does not generate an inconsistency. The \textit{ub-grounded} semantics blocks undecidedness \textit{as earlier as possible} outside a cyclic strongly connected component and it lets undecidedness propagate inside a cyclic strongly connected component. 

The \textit{ub-preferred} semantics and the \textit{weakly preferred} semantics use the undecidedness blocking condition respectively \textit{as much as possible} and \textit{as earlier and as much as possible}. 

The ground-based algorithm suggests that the generation of Dung's \textit{complete} semantics can also be revisited as a special case of \textit{undecidedness blocking} semantics with the extra condition of admissibility.
The admissibility condition requires the accepted arguments to be able to collectively defend themselves by labelling \texttt{out} their attackers. 
This means that in Dung's \textit{complete} semantics not only undecidedness is blocked by using the \textit{ub-condition}, but it is also \textit{resolved}. The conflicts between arguments that created the undecided situation are resolved by promoting some of them to the label \texttt{in}, so that none of these arguments is attacked by an undecided argument anymore in the resulting labelling. Since the undecidedness is solved, in the resulting labelling there is no trace that it was blocked.

The diagram in Figure 8 represents the relations among Dung’s \textit{complete} semantics and the \textit{weakly complete} semantics introduced in this paper.

\begin{figure}[h]
\begin{center}
\includegraphics[scale=0.4]{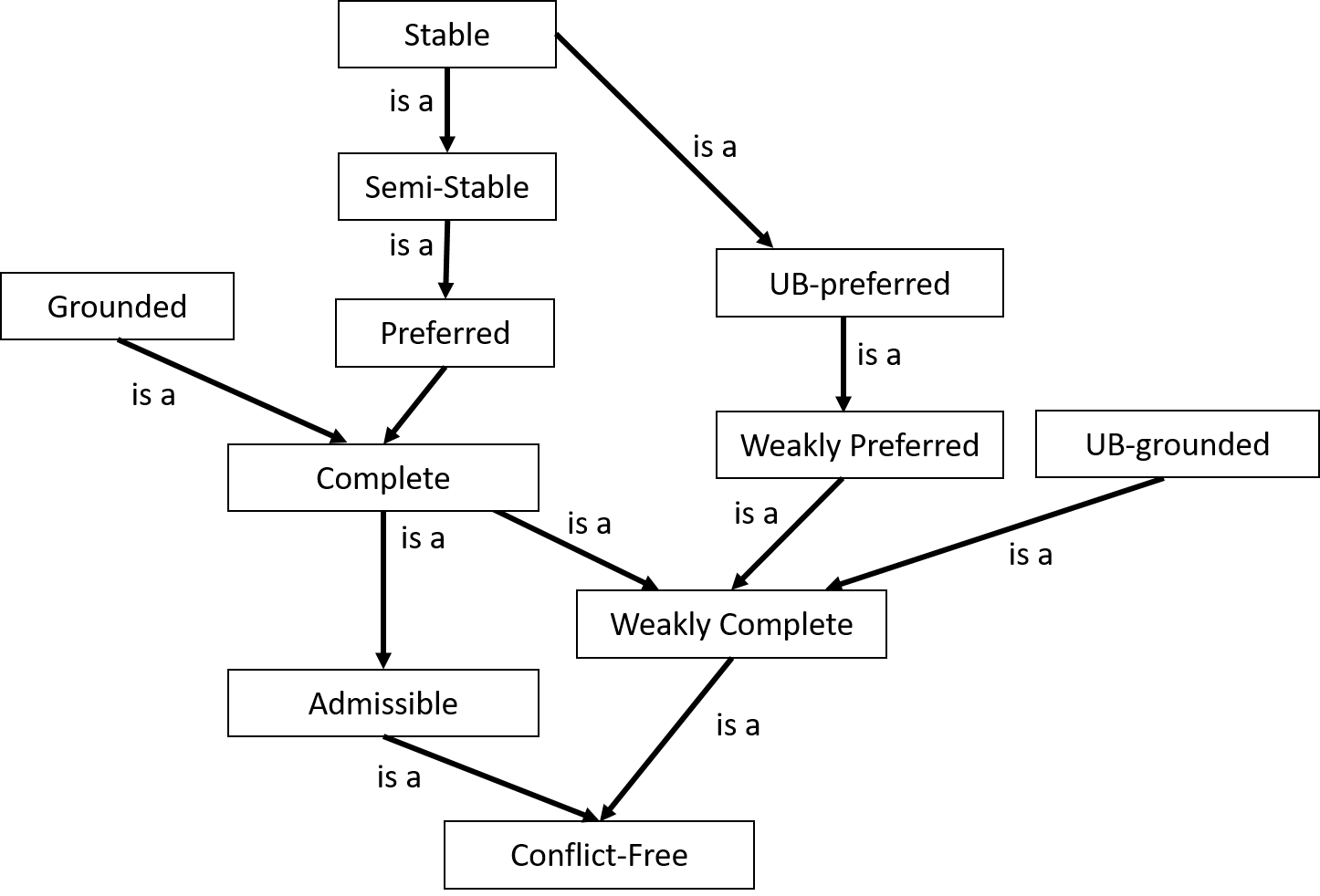}
\caption{Hierarchy of \textit{weakly complete} semantics}
\label{fig1}
\vspace{-5mm}
\end{center}
\end{figure}

\subsection{Weakly Complete versus Baumann-Brewka-Ulbricht semantics}
Baumann et al. \cite{bbu1} proposed a solution to the problem of self-defeating attackers, defining a family of semantics (here called $BBU$ semantics) based on a weaker notion of Dung’s admissibility. 
While the definition 3.1 of our \textit{weakly complete} semantics requires a small modification of Dung's \textit{complete} labelling, the definition of $BBU$ semantics require some preliminary concepts.
Given an argumentation framework $AF=\langle Ar,\mathcal{R} \rangle$ and a set $E \in Ar$, we define the set $AF^E$, called the \textit{E-reduct} of $AF$ as its restriction $AF_{ \downarrow E^{*}}$, where $E^{*}=A \setminus (E \cup E^{+})$, that is the restriction containing all the arguments except the arguments in $E$ and the ones attacked by $E$.  

A \textit{weakly admissible} set $S$ is a set that is required to defend itself only from attacks of arguments that are \textit{weakly admissible} in $AF^S$. Formally:

\vspace{3mm}

\noindent \textbf{Definition 6.3} \textit{Given the framework $AF=\langle Ar,\mathcal{R} \rangle$, the set of weakly admissible sets of $AF$ is denoted $ad^w(AF)$ and defined by $E \in ad^w(AF)$ iff $E$ is conflict-free and for every attacker $y$ of $E$ we have $y \in \bigcup ad^w(AF^E)$. }.
\vspace{3mm}

\noindent Based on the notion of \textit{weakly admissible} sets, the authors defined the concept of weak defence, so that every conflict-free set is \textit{weakly admissible} if and only if it weakly defends itself.

\vspace{3mm}
\noindent \textbf{Definition 6.4} \textit{Given the framework $AF=\langle Ar,\mathcal{R} \rangle$, a set $E \subseteq Ar$ weakly defends a set $X \subseteq Ar$ whenever, for every attacker $y$ of $X$, either $E$ attacks $y$, or $y \not \in \bigcup ad^w(AF^E)$, $y \not \in E$ and $X \subseteq X^{'} \in ad^w(AF^E)$}.
\vspace{3mm}

The $BBU$ complete, preferred and grounded semantics are defined as follows:

\vspace{3mm}
\noindent \textbf{Definition 6.5} \textit{Given $AF=\langle Ar,\mathcal{R} \rangle$ and $E \subseteq Ar$, we say that $E$ is:
\begin{itemize}
    \item a BBU complete extension of $AF$ $(E \in co_{bbu}(AF))$ iff $E \in ad^w(AF)$ and for every $X$ such that $E \subseteq X$ that is weakly defended by $E$, we have $X \subseteq E$.
    \item a BBU preferred extension of $AF$ $(E \in pr_{bbu}(AF))$ iff $E$ is maximal w.r.t. set inclusion in $ad^w(F)$.
    \item a BBU preferred extension of $AF$ $(E \in gr_{bbu}(AF))$ iff $E$ is minimal w.r.t. set inclusion in $co^w(F)$.
\end{itemize}
}
\vspace{3mm}

We start our comparison by observing how our semantics and the $BBU$ semantics do not coincide.
Let’s consider the floating assignment example in figure 1A. According to the definition of \textit{weakly admissible} set, the set $\{a\}$ is weakly admissible ($F^{a}= \emptyset$) and so is $\{b\}$. This implies that $c$ is not in any \textit{weakly admissible} set since $a$ and $b$ are. The $BBU$ \textit{complete} extensions are $\{\{a\}, \{b\}, \emptyset \}$, coinciding with Dung’s semantics. 
Regarding graphs $B$ and $C$ in figure 1, $b$ is accepted since a self-attacking argument (graph 1B) or an odd-length cycle not externally attacked (1C) are not \textit{weakly admissible} sets.

Our \textit{weakly complete} semantics generate the same labelling for graph $B$ and $C$ of figure 1, while in graph 1A it adds the additional labelling where $c$ is accepted and $a$ and $b$ are undecided.
In graph $1A$, our semantics allow for the interpretation where the attacks to $c$ are from two arguments that are conflicting and therefore not \textit{strong enough} to defeat $c$. This is certainly the case if we take a \textit{grounded semantics} stance on the two rebutting arguments $a$ and $c$. 
Indeed graphs $A,B$ and $C$ of figure 1 could be interpreted as instances of the same situation where argument $b$ is receiving attacks from conflicting arguments. The only difference is that in graph $A$ argument $b$ is attacked by an even-length cycle. Indeed, the \textit{preferred semantics} interpretation of graph $A$ is a valid (and it is indeed included in our \textit{weakly complete} labelling), but not the only reasonable one. We believe the extra labelling added by \textit{weakly complete} semantics shows how such semantics are able to transfer the sceptical stance proper of \textit{grounded} semantics to the non-admissible case, while the non-admissible $BBU$ version of \textit{grounded} semantics does not.

\begin{figure}[h]
\begin{center}
\includegraphics[scale=0.6]{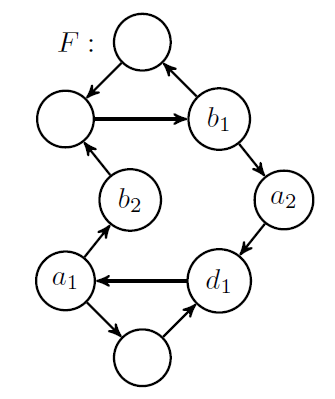}
\caption{A graph with two BBU weakly grounded extensions $  \{a_1,b_1\}, \{a_2,b_2\}$. From \cite{bbu1} }
\label{fig9}
\vspace{-5mm}
\end{center}
\end{figure}

\noindent The principle-based analysis of the two families of semantics reveals some important differences. The results for $BBU$ semantics are taken from \cite{bub_principle}. We report our main observations:
\begin{itemize}
    \item \textit{Grounded} semantics. The $BBU$ version of \textit{grounded} semantics strongly deviates from Dung’s counterpart. There could be multiple $BBU$ \textit{grounded} extensions of a framework, therefore losing the single-status property, and it is possible that none of those extensions is Dung’s \textit{grounded} semantics. For instance, the graph of figure 9 has two \textit{grounded} $BBU$ extension: $  \{a_1,b_1\}, \{a_2,b_2\}$, while Dung's  \textit{grounded} semantics is empty. On the contrary, both our  \textit{weakly grounded} and \textit{ub-grounded} semantics are single-status, they always exist and they can be computed in polynomial computational time. Our semantics retain the scepticism of Dung’s grounded semantics  in the way cycles are treated, while the $BBU$ version of \textit{grounded} semantics loses much of the \textit{sceptical stance} of its Dung's counterparts. 
\item \textit{Directionality}. $BBU$ \textit{complete} and \textit{grounded} semantics do not satisfy \textit{directionality} and no result has been proved for \textit{preferred} $BBU$ semantics. On the contrary, \textit{weakly complete} and \textit{ub-semantics} do satisfy \textit{directionality}.
\item \textit{Abstention}. None of the $BBU$ semantics satisfy \textit{abstention}. \textit{Abstention} is an interesting property satisfied by Dung’s \textit{complete} semantics. If an argument is labelled \texttt{out} in one valid labelling and \texttt{in} in another, the argument is \textit{disputed}, and a semantics satisfying \textit{abstention} also provides at least a third labelling where such an argument is  \texttt{undec}. 
Our \textit{weakly complete} semantics satisfy \textit{abstention}. Our \textit{grounded} and \textit{ub-grounded} trivially satisfy it, while the \textit{preferred} does not. Therefore, they indeed behave like Dung’s counterparts. 
\item \textit{SCC-Decomposition}. Negative results are proved for $BBU$ \textit{complete} and \textit{grounded} semantics, and no proof is provided for \textit{preferred} semantics. Our \textit{ub-preferred} and \textit{ub-grounded} semantics are \textit{SCC-recursive}, and therefore they are guaranteed to satisfy \textit{SCC-Decomposition}. According to proposition 55 in \cite{baroni2007principle},
\textit{weakly preferred} semantics is not \textit{SCC-decomposable} since it has a cardinality of extensions greater than zero and it does not satisfy \textit{directionality}.
\item \textit{Set inclusion}. $BBU$ \textit{weakly complete} semantics do not include Dung’s \textit{complete} semantics, while Dung’s \textit{complete} labellings are a subset of our \textit{weakly complete} semantics labellings.
\end{itemize}

In general, the above principle-based analysis shows how our semantics satisfy all the principles satisfied by $BBU$ semantics, but they also satisfy other principles like \textit{directionality}, \textit{SCC-decomposition} and \textit{abstention} that are satisfied by Dung’s counterparts. Our semantics results indeed closer to the principles underlying Dung’s \textit{complete} semantics; \textit{weakly complete} and \textit{ub-grounded} semantics exactly satisfy the same principles of Dung’s counterparts. Based on this, we believe that \textit{weakly complete} semantics represent a more faithful extension of Dung’ semantics to the non-admissible case and a more faithful solution to the problem of self-defeating attacking arguments.

\subsection{Ambiguity Blocking and Undecidedness Blocking}
The introduction of \textit{weakly complete} semantics was motivated by the idea of introducing \textit{ambiguity blocking} mechanisms in abstract argumentation. 
So far our semantics have been presented as \textit{undecidedness blocking} semantics in a self-contained way, without references to \textit{ambiguity blocking} mechanisms present in other non-monotonic formalisms.
Nevertheless, \textit{undecidedness blocking} is indeed inspired by the mechanism of \textit{ambiguity blocking} and the willingness to solve situations like the floating assignment example in the way \textit{ambiguity blocking} semantics do, that is accepting the attacked argument on the basis that the two attackers are conflicting. 

In this section we explore the relation between \textit{ambiguity blocking} and \textit{undecidedness blocking}. The relation is a loose one; the two concepts do not coincide but nevertheless they share strong conceptual similarities. We claim how undecidedness is a broader concept able to model ambiguity but not limited to it.

We first need to specify what we mean by ambiguity. Here we refer to the concept of ambiguity as defined in defeasible logic. 
Defeasible logic \cite{nute2001defeasible} is a rule-based non-monotonic formalism. Rules can be strict ($A \rightarrow B$), representing monotonic rules where $B$ follows always if $A$ is true or defeasible rule ($A \Rightarrow B$), meaning that from $A$ defeasibly follows $B$. Defeasible rules represent default positions that are assumed valid unless contrary evidence is provided. Rules with empty body represent facts. $\rightarrow B$ is an indisputable fact, $\Rightarrow B$ is a defeasible fact.
An acyclic \textit{superiority relation} is used to define priorities among rules, that is, where one rule may override the conclusion of another rule. 
Conclusions can be classified as \textit{definite} (also called \textit{strict}) or \textit{defeasible}. A conclusion can be therefore provable (or not provable) strictly or defeasibly.
A rule is applicable if its antecedents are defeasible probable. A rules is not applicable if it has been proved that at least one antecedent of the rule is not defeasible provable. Strict conclusions are obtained by forward chaining of strict rules, while a defeasible conclusion $A$ can be derived if there is an applicable strict or defeasible rule with conclusion $A$, and either (1) $\lnot A$ is not definitely provable and each rules concluding $\lnot A$ has been proved to be not applicable, or (2) every applicable rule for $\lnot A$ is weaker (according to the superiority relation) than an applicable strict or defeasible rule for $A$.

A \textit{defeasible theory} is a set of rules and a superiority relation. 
A literal $a$ is \textit{ambiguous} if there is a valid chain of reasoning concluding $a$ and another concluding $\lnot a$ and the superiority relation cannot resolve such conflict. The \textit{ambiguity blocking} mechanism of defeasible logic semantics is that both $a$ and $\lnot a$ are refuted and their ambiguity is not propagated to other conflicting literals, as shown in the following example.
\vspace{3mm}

\noindent \textbf{Example 6.1}. Let us consider the following defeasible theory:
\begin{equation}
\mathcal{D}=\{\Rightarrow a, \; \Rightarrow \lnot a, \; \lnot a \Rightarrow b, \; \Rightarrow \lnot b \}
\end{equation}

In the \textit{ambiguity blocking} defeasible logics semantics only $\lnot b$ can be defeasibly proved. Indeed, since literal $a$ and $\lnot a$ are ambiguous (we have the two conflicting facts $\Rightarrow a$ and $\Rightarrow \lnot a$), they are both refuted and therefore the rule $\lnot a \Rightarrow b$ is not applicable, and $\lnot b$ is proved. Using an \textit{ambiguity propagating} semantics none of the literals could be proved and both $a$ and $b$ are ambiguous.
\vspace{3mm}

\noindent \textbf{Example 6.2}. The following theory models a conflict similar to the floating assignment situation:
\begin{equation}
\mathcal{D}=\{\Rightarrow e, \; \Rightarrow \lnot e, \;  e \Rightarrow g, \; \lnot e \Rightarrow  g \}
\end{equation}
where the literal $g$ means \textit{guilty} and $e$ and $\lnot e$ are the conflicting evidence used to accuse the defendant. Using \textit{ambiguity blocking} $g$ cannot be proved since $e$ is ambiguous.
\vspace{3mm}

While ambiguity is well defined in \textit{defeasible logic} ($DL$), in Dung’s abstract argumentation the concept of \textit{ambiguous} argument is not even defined. In \cite{governatori2004argumentation} Governatori et al. referred to Dung’s \textit{grounded} semantics as an \textit{ambiguity propagating} semantics, but their study is in the context of defeasible logic, not abstract argumentation, and a definition of ambiguous argument in Dung's framework is outside the scope of their work.
Our intuition was to model ambiguity with undecidedness. What is propagated by \textit{grounded} semantics is indeed the undecided status of arguments.

\textit{Ambiguity blocking} and \textit{undecidedness blocking} have some strong similarities. First of all, ambiguous literals and undecided arguments both signal unresolved conflicts. In the first case the superiority relation does not help to resolve the conflict, while in the second the postulates of the semantics do not provide a definitive reason to accept or reject the conflicting arguments.
Second, both ambiguity and undecidedness are blocked by rejecting unresolved conflicts. The ambiguity blocking behaviour is to reject the ambiguous literals and all the rules containing ambiguous literals in their premises. Literals that were in conflict with such rejected literals become provable. Undecidedness blocking semantics do indeed mimic this behaviour. The mechanism blocks attacks from undecided arguments so that attacked arguments become now acceptable.

There are also notable differences. Conflicts in $DL$ are represented by complementary literals, defining a symmetric conflict relation. The superiority relation of defeasible logic is by definition acyclic, meaning that the clash of complementary literals is the only form of “cyclic” conflict in $DL$.
Two symmetrical conflicting arguments are most likely to be modelled in abstract argumentation by a set of two rebutting arguments, but it is not guaranteed that those arguments will generate undecidedness.
This indeed depends on the semantics used: undecidedness is generated if we are using \textit{grounded} semantics, but not if we are using \textit{preferred} semantics.
Linking ambiguity with undecidedness is therefore semantics-dependant, since what is undecided depends on the stance of the semantics employed, hence the different families of \textit{undecidedness-blocking} semantics presented in this paper. 

Moreover, undecidedness can arise in situations where there is no ambiguity. Paradoxical situations like a cycle of three arguments with unidirectional attacks also generate an undecided situation in abstract argumentation, but they are not possible in defeasible logic since the superiority relation is required to be acyclic. Therefore undecidedness is a much general concept signalling unresolved conflicts, including dilemma or paradoxical situations that might not have anything to do with ambiguity. Indeed, if we consider the meaning of the English word \textit{ambiguous} (that is \textit{"a situation open to multiple interpretations"}), is clearly not the same as being undecided. For instance, in a paradox there is no ambiguity since we do not have multiple potentially valid interpretations but rather all the interpretations are contradictory. These observations suggest how ambiguity is a special case of the more general concept of undecidedness.
Therefore, \textit{undecidedness blocking} semantics could be used to generalize \textit{ambiguity blocking} semantics to the case of cyclic conflict relations among arguments and to model less sceptical multi-status \textit{ambiguity blocking} semantics.

A future research direction is the formal investigation of the relation between defeasible logic \textit{ambiguity blocking} and our \textit{undecidedness blocking} semantics. Our hypothesis is that the \textit{ub-grounded} semantics can be used to instantiate \textit{ambiguity blocking} defeasible logic semantics in the same way \textit{grounded} semantics was used to instantiate \textit{ambiguity propagation} defeasible logic semantics in \cite{lam2016aspic+}. Since $DL$ is a rule-based structured formalism while Dung’s framework is abstract, a formal analysis of the possibility of using our semantics to instantiate \textit{ambiguity blocking} would require to use structured argumentation frameworks such as $ASPIC+$.

\section{Related Works}
One of the main reason to introduce \textit{weakly complete} semantics was to translate \textit{ambiguity blocking} into abstract argumentation.
The work in \cite{governatori2004argumentation} represents our main reference regarding the link between \textit{ambiguity blocking} defeasible logic and abstract argumentation. We want to stress the difference between the two works: while in \cite{governatori2004argumentation} authors translate Dung' semantics into \textit{ambiguity propagating} defeasible logic, in this paper we went the opposite direction and we translated the notion of \textit{ambiguity blocking} into abstract argumentation.

This paper extends our preliminary works \cite{dondio2018proposal,dondio2019beyond}, where we proposed an abstract semantics to model the principle of \textit{beyond reasonable doubt}. The resulting semantics was a first attempt to define an \textit{undecidedness blocking} semantics. The semantics proposed is a subset of \textit{weakly complete} semantics, the \textit{ub-grounded} semantics is defined in \cite{dondio2019beyond} but its links to defeasible logics and \textit{ambiguity blocking} semantics are not studied.
A Dung-like version of \textit{ambiguity blocking} has been investigated in the context of the instantiation of the Carneades argumentation system \cite{gordon2006carneades} into the Dung-based structured argumentation system ASPIC+ \cite{modgil2014aspic+}.
For instance, in \cite{van2012relating} a translation mechanism is proposed to model the Carneades argumentation systems into an ASPIC+ argumentation system. The relevance to our work lies in the fact that the Carneades argumentation system is \textit{ambiguity blocking} while ASPIC+, by relying on Dung' semantics, is \textit{ambiguity propagating}. Therefore a translation from Carneades to ASPIC+ has to deal with the problem of modelling \textit{ambiguity blocking} in a Dung-like system. The authors state how this was the \textit{"the main difficulty"} of the translation process. Instead of introducing a new Dunganian semantics, the authors solved the problem by introducing additional argument nodes, allowing for an explicit representation of applicability and acceptability of rules. In this translation, a couple undercutter defeaters (unidirectional attacks) is added for each contradictory literals to refute both of them and replicate the ambiguity-blocking behaviour. The authors do not propose a new ambiguity-blocking Dunganian semantics, but in the context of a structured argumentation system they mimic the ambiguity blocking behaviour using Dung’ semantics and additional nodes.

This approach is implicitly questioned in \cite{lam2016aspic+}, where the relation between Defeasible logic and ASPIC+ is investigated. The authors show how ASPIC+ using \textit{grounded} semantics is equivalent to the \textit{ambiguity propagation} version of defeasible logic semantics, and they consider how the \textit{ambiguity blocking} $DL$ semantics can be instantiated in ASPIC+. The authors conclude how such a translation could result problematic; the $DL$ with \textit{ambiguity blocking} would require to introduce a second \textit{“attack”} relation on arguments with a ripple down effects on the ASPIC+ definitions setting the various statuses of the argument.
Therefore, an \textit{ambiguity blocking} abstract semantics has not been developed, and in the context of structured argumentation there are works proposing non-trivial translations requiring the addition of new nodes and meta-concepts.

Similar conclusions have been reached by works on \textit{standard of proof} and legal reasoning applied to abstract argumentation, since legal reasoning is often \textit{ambiguity blocking}.

In particular, the standard of proof \textit{beyond reasonable doubt} is responsible for the \textit{ambiguity blocking} mechanism of legal reasoning, since evidence that are not \textit{beyond reasonable doubt} are deemed not sufficiently credible and therefore \textit{blocked}.
Standards of proof have been extensively study in argumentation theory \cite{gordon2009proof}, but only few studies are relevant to abstract argumentation. In the context of structured argumentation, we mention the work of Prakken \cite{prakken2011modelling} on modelling standards of proof, and the modification of the Carneades framework \cite{gordon2006carneades} to accommodate various standards of proof. 
Regarding abstract argumentation, the most explicit study about standards of proof is \cite{atkinson2007argumentation}. Here, the authors consider how each Dung’s semantics has a different level of cautiousness that is mapped to a corresponding legal standard of proof. Only initial arguments are beyond doubt, but they consider the sceptically preferred justification a beyond reasonable doubt position. In the floating assignment example (shown in Figure 1), the authors recognize the two attackers as doubtful, but they consider the sceptically preferred rejection of the attacked argument beyond reasonable doubt. It could be noticed that this position is failing to acknowledge that, if each of the attackers are considered doubtful, their effect cannot be (at last in all the situations) beyond doubt.
Brewka et al. \cite{brewka2010carneades} also criticise \cite{atkinson2007argumentation} since they doubt the fact that various Dung’s semantics can capture the intuitive meaning of legal standards of proof (detailed discussion in here \cite{gordon2009proof}). In the case of beyond reasonable doubt, we agree with Brewka, \textit{complete} Dung’s semantics are not adequate to model this principle. 

Prakken has analysed the floating assignment and its link to standards of proof in his work \cite{prakken2002intuitions}, where he responds to objections advanced by Horty in \cite{horty2001argument}. Prakken underlines that, in many problematic situations including the floating assignment, there could be hidden assumptions about the specific problem which, if made explicit, are nothing but extra information that defeat the defeasible inference. In the case of the floating assignment, Prakken agrees that if beyond reasonable doubt is our standard of proof - like in a criminal case where there are two conflicting testimonies - we should not conclude that the accused is guilty. 

In his presentation of \textit{semi-stable} semantics, Caminada \cite{caminada2012semi} also provides another example of the logical assumptions that could be hidden beyond the treatment of the floating assignment. In particular, he observes how the preferred semantics solution is based on the assumption that we know with certainty that one of the two attacking arguments is valid, since in this case we do not need to know which one is valid in order to safely discard the third argument.
However, the above observations do not mean that argumentation semantics are somehow invalid. In the case of conflicting testimonies, as already showed by Pollock \cite{pollock1995cognitive}, the situation could be correctly modelled by making some hidden assumptions explicit and adding extra arguments to model such assumptions. In the conflicting testimonies, the fact that two witnesses contradict each other is a reason to add an argument undercutting the credibility of both. Instead of adding arguments interacting with existing arguments, in this work we have tackled the problem by embedding assumptions directly in a novel abstract argumentation semantics.

\section{Conclusions and Future Works}
In this paper we have explored a new family of semantics called \textit{weakly complete} semantics. Unlike Dung' semantics, in \textit{weakly complete} semantics the undecided label can be blocked by the postulates of the semantics rather than being propagated from an undecided attacking argument to an otherwise accepted attacked argument.  

The new semantics are conflict-free, non-admissible (in Dung’s sense), but employing a more relaxed defence-based notion of admissibility; they allow reinstatement, they generate extensions that are super sets of \textit{grounded} semantics and they retain the majority of properties of \textit{complete} semantics. The semantics can also provide a solution to the 25-year old problem of self-defeating attackers.
We have studied the properties of these new semantics and we have provided an algorithm to compute them. The algorithm suggests an interpretation of the various \textit{weakly complete} and \textit{complete} semantics as different strategies to reduce undecidedness.
We have identified families of \textit{weakly complete} semantics satisfying specific constraints on the set of acceptable arguments and employing different \textit{undecidedness blocking} strategies.
Regarding computational complexity, the main result is that the credulous acceptance problem for \textit{weakly complete} semantics can be solved in polynomial time. 

We have performed a principle-based comparison between our semantics and the \textit{weakly admissible} semantics recently proposed by Baumann et al. 
The analysis shows how our semantics satisfy  a  number  of  principles  satisfied  by  Dung’s complete semantics  but not by Baumann et al.  semantics, including \textit{directionality, abstention, SCC-decomposability} and \textit{cardinality} of extensions, making them a more faithful translation of Dung’ semantics to the non-admissible case.

Future research directions include the optimization of the computational algorithm proposed and the formal investigation of how \textit{ub-grounded} semantics can be used in Dung-based structured frameworks like $ASPIC+$ to instantiate \textit{ambiguity blocking} defeasible logic semantics. It also includes the comparison of the newly proposed weakly complete semantics empirically \cite{rizzo2018investigation} and its comparison against other semantics in order to evaluate its inferential capacity.
\label{S:2}






\bibliographystyle{elsarticle-num-names}
\bibliography{wk}

\begin{thebibliography}{29}
\expandafter\ifx\csname natexlab\endcsname\relax\def\natexlab#1{#1}\fi
\providecommand{\url}[1]{\texttt{#1}}
\providecommand{\href}[2]{#2}
\providecommand{\path}[1]{#1}
\providecommand{\DOIprefix}{doi:}
\providecommand{\ArXivprefix}{arXiv:}
\providecommand{\URLprefix}{URL: }
\providecommand{\Pubmedprefix}{pmid:}
\providecommand{\doi}[1]{\href{http://dx.doi.org/#1}{\path{#1}}}
\providecommand{\Pubmed}[1]{\href{pmid:#1}{\path{#1}}}
\providecommand{\bibinfo}[2]{#2}
\ifx\xfnm\relax \def\xfnm[#1]{\unskip,\space#1}\fi
\bibitem[{Dung(1995)}]{dung}
\bibinfo{author}{P.~M. Dung},
\newblock \bibinfo{title}{On the acceptability of arguments and its fundamental
  role in nonmonotonic reasoning, logic programming and n-person games},
\newblock \bibinfo{journal}{Artificial intelligence} \bibinfo{volume}{77}
  (\bibinfo{year}{1995}) \bibinfo{pages}{321--357}.
\bibitem[{Caminada and Gabbay(2009)}]{caminada}
\bibinfo{author}{M.~W. Caminada}, \bibinfo{author}{D.~M. Gabbay},
\newblock \bibinfo{title}{A logical account of formal argumentation},
\newblock \bibinfo{journal}{Studia Logica} \bibinfo{volume}{93}
  (\bibinfo{year}{2009}) \bibinfo{pages}{109--145}.
\bibitem[{Baumann et~al.(2020)Baumann, Brewka, and Ulbricht}]{bbu1}
\bibinfo{author}{R.~Baumann}, \bibinfo{author}{G.~Brewka},
  \bibinfo{author}{M.~Ulbricht},
\newblock \bibinfo{title}{Revisiting the foundations of abstract
  argumentation--semantics based on weak admissibility and weak defense},
\newblock in: \bibinfo{booktitle}{Proceedings of the AAAI Conference on
  Artificial Intelligence}, volume~\bibinfo{volume}{34}, \bibinfo{year}{2020},
  pp. \bibinfo{pages}{2742--2749}.
\bibitem[{Calegari et~al.(2019)Calegari, Contissa, Lagioia, Omicini, and
  Sartor}]{calegari2019defeasible}
\bibinfo{author}{R.~Calegari}, \bibinfo{author}{G.~Contissa},
  \bibinfo{author}{F.~Lagioia}, \bibinfo{author}{A.~Omicini},
  \bibinfo{author}{G.~Sartor},
\newblock \bibinfo{title}{Defeasible systems in legal reasoning: A comparative
  assessment.},
\newblock in: \bibinfo{booktitle}{JURIX}, \bibinfo{year}{2019}, pp.
  \bibinfo{pages}{169--174}.
\bibitem[{van Gijzel and Prakken(2012)}]{van2012relating}
\bibinfo{author}{B.~van Gijzel}, \bibinfo{author}{H.~Prakken},
\newblock \bibinfo{title}{Relating carneades with abstract argumentation via
  the aspic+ framework for structured argumentation},
\newblock \bibinfo{journal}{Argument \& Computation} \bibinfo{volume}{3}
  (\bibinfo{year}{2012}) \bibinfo{pages}{21--47}.
\bibitem[{Governatori et~al.(2004)Governatori, Maher, Antoniou, and
  Billington}]{governatori2004argumentation}
\bibinfo{author}{G.~Governatori}, \bibinfo{author}{M.~J. Maher},
  \bibinfo{author}{G.~Antoniou}, \bibinfo{author}{D.~Billington},
\newblock \bibinfo{title}{Argumentation semantics for defeasible logic},
\newblock \bibinfo{journal}{Journal of Logic and Computation}
  \bibinfo{volume}{14} (\bibinfo{year}{2004}) \bibinfo{pages}{675--702}.
\bibitem[{Modgil and Prakken(2014)}]{modgil2014aspic+}
\bibinfo{author}{S.~Modgil}, \bibinfo{author}{H.~Prakken},
\newblock \bibinfo{title}{The aspic+ framework for structured argumentation: a
  tutorial},
\newblock \bibinfo{journal}{Argument \& Computation} \bibinfo{volume}{5}
  (\bibinfo{year}{2014}) \bibinfo{pages}{31--62}.
\bibitem[{Lam et~al.(2016)Lam, Governatori, and Riveret}]{lam2016aspic+}
\bibinfo{author}{H.-P. Lam}, \bibinfo{author}{G.~Governatori},
  \bibinfo{author}{R.~Riveret},
\newblock \bibinfo{title}{On aspic+ and defeasible logic.},
\newblock in: \bibinfo{booktitle}{COMMA}, \bibinfo{year}{2016}, pp.
  \bibinfo{pages}{359--370}.
\bibitem[{Baumann et~al.(2020)Baumann, Brewka, and
  Ulbricht}]{baumann2020comparing}
\bibinfo{author}{R.~Baumann}, \bibinfo{author}{G.~Brewka},
  \bibinfo{author}{M.~Ulbricht},
\newblock \bibinfo{title}{Comparing weak admissibility semantics to their
  dung-style counterparts--reduct, modularization, and strong equivalence in
  abstract argumentation},
\newblock in: \bibinfo{booktitle}{Proceedings of the International Conference
  on Principles of Knowledge Representation and Reasoning},
  volume~\bibinfo{volume}{17}, \bibinfo{year}{2020}, pp.
  \bibinfo{pages}{79--88}.
\bibitem[{Dondio and Longo(2019)}]{dondio2019beyond}
\bibinfo{author}{P.~Dondio}, \bibinfo{author}{L.~Longo},
\newblock \bibinfo{title}{Beyond reasonable doubt: A proposal for undecidedness
  blocking in abstract argumentation},
\newblock \bibinfo{journal}{Intelligenza Artificiale} \bibinfo{volume}{13}
  (\bibinfo{year}{2019}) \bibinfo{pages}{123--135}.
\bibitem[{Dondio and Longo(2018)}]{dondio2018proposal}
\bibinfo{author}{P.~Dondio}, \bibinfo{author}{L.~Longo},
\newblock \bibinfo{title}{A proposal to embed the in dubio pro reo principle
  into abstract argumentation semantics based on topological ordering and
  undecidedness propagation.},
\newblock in: \bibinfo{booktitle}{AI$^3$@ AI* IA}, \bibinfo{year}{2018}, pp.
  \bibinfo{pages}{42--56}.
\bibitem[{Dondio(2019)}]{dondio_tech}
\bibinfo{author}{P.~Dondio},
\newblock \bibinfo{title}{Weakly-admissible semantics and the problem of
  ambiguity blocking in abstract argumentation},
\newblock in: \bibinfo{booktitle}{Technical Report, Technological University
  Dublin}, \bibinfo{year}{2019}, pp. \bibinfo{pages}{1--8}.
\bibitem[{Baroni et~al.(2005)Baroni, Giacomin, and Guida}]{baroni2005scc}
\bibinfo{author}{P.~Baroni}, \bibinfo{author}{M.~Giacomin},
  \bibinfo{author}{G.~Guida},
\newblock \bibinfo{title}{Scc-recursiveness: a general schema for argumentation
  semantics},
\newblock \bibinfo{journal}{Artificial Intelligence} \bibinfo{volume}{168}
  (\bibinfo{year}{2005}) \bibinfo{pages}{162--210}.
\bibitem[{Baroni et~al.(2011)Baroni, Caminada, and
  Giacomin}]{baroni2011introduction}
\bibinfo{author}{P.~Baroni}, \bibinfo{author}{M.~Caminada},
  \bibinfo{author}{M.~Giacomin},
\newblock \bibinfo{title}{An introduction to argumentation semantics},
\newblock \bibinfo{journal}{The Knowledge Engineering Review}
  \bibinfo{volume}{26} (\bibinfo{year}{2011}) \bibinfo{pages}{365--410}.
\bibitem[{van~der Torre and Vesic(2017)}]{van2017principle}
\bibinfo{author}{L.~van~der Torre}, \bibinfo{author}{S.~Vesic},
\newblock \bibinfo{title}{The principle-based approach to abstract
  argumentation semantics},
\newblock \bibinfo{journal}{IfCoLog Journal of Logics and Their Applications}
  (\bibinfo{year}{2017}).
\bibitem[{Jeremie~Dauphin(2020)}]{bub_principle}
\bibinfo{author}{L.~V. D.~T. Jeremie~Dauphin, Tjitze~Rienstra},
\newblock \bibinfo{title}{A principle-based analysis of weakly admissible
  semantics},
\newblock \bibinfo{journal}{Computational Models of Argument: Proceedings of
  COMMA 2020} \bibinfo{volume}{326} (\bibinfo{year}{2020})
  \bibinfo{pages}{167}.
\bibitem[{Baroni and Giacomin(2007)}]{baroni2007principle}
\bibinfo{author}{P.~Baroni}, \bibinfo{author}{M.~Giacomin},
\newblock \bibinfo{title}{On principle-based evaluation of extension-based
  argumentation semantics},
\newblock \bibinfo{journal}{Artificial Intelligence} \bibinfo{volume}{171}
  (\bibinfo{year}{2007}) \bibinfo{pages}{675--700}.
\bibitem[{Pollock(2001)}]{pollock2001defeasible}
\bibinfo{author}{J.~L. Pollock},
\newblock \bibinfo{title}{Defeasible reasoning with variable degrees of
  justification},
\newblock \bibinfo{journal}{Artificial intelligence} \bibinfo{volume}{133}
  (\bibinfo{year}{2001}) \bibinfo{pages}{233--282}.
\bibitem[{Nute(2001)}]{nute2001defeasible}
\bibinfo{author}{D.~Nute},
\newblock \bibinfo{title}{Defeasible logic},
\newblock in: \bibinfo{booktitle}{International Conference on Applications of
  Prolog}, \bibinfo{organization}{Springer}, \bibinfo{year}{2001}, pp.
  \bibinfo{pages}{151--169}.
\bibitem[{Gordon and Walton(2006)}]{gordon2006carneades}
\bibinfo{author}{T.~F. Gordon}, \bibinfo{author}{D.~Walton},
\newblock \bibinfo{title}{The carneades argumentation},
\newblock in: \bibinfo{booktitle}{Computational models of argument: Proceedings
  of COMMA 2006}, volume \bibinfo{volume}{144}, \bibinfo{organization}{IOS
  Press}, \bibinfo{year}{2006}, p. \bibinfo{pages}{195}.
\bibitem[{Gordon and Walton(2009)}]{gordon2009proof}
\bibinfo{author}{T.~F. Gordon}, \bibinfo{author}{D.~Walton},
\newblock \bibinfo{title}{Proof burdens and standards},
\newblock in: \bibinfo{booktitle}{Argumentation in artificial intelligence},
  \bibinfo{publisher}{Springer}, \bibinfo{year}{2009}, pp.
  \bibinfo{pages}{239--258}.
\bibitem[{Prakken and Sartor(2011)}]{prakken2011modelling}
\bibinfo{author}{H.~Prakken}, \bibinfo{author}{G.~Sartor},
\newblock \bibinfo{title}{On modelling burdens and standards of proof in
  structured argumentation.},
\newblock in: \bibinfo{booktitle}{JURIX}, \bibinfo{organization}{Citeseer},
  \bibinfo{year}{2011}, pp. \bibinfo{pages}{83--92}.
\bibitem[{Atkinson and Bench-Capon(2007)}]{atkinson2007argumentation}
\bibinfo{author}{K.~Atkinson}, \bibinfo{author}{T.~Bench-Capon},
\newblock \bibinfo{title}{Argumentation and standards of proof},
\newblock in: \bibinfo{booktitle}{Proceedings of the 11th international
  conference on Artificial intelligence and law}, \bibinfo{year}{2007}, pp.
  \bibinfo{pages}{107--116}.
\bibitem[{Brewka and Gordon(2010)}]{brewka2010carneades}
\bibinfo{author}{G.~Brewka}, \bibinfo{author}{T.~F. Gordon},
\newblock \bibinfo{title}{Carneades and abstract dialectical frameworks: A
  reconstruction},
\newblock in: \bibinfo{booktitle}{Proceedings of the 2010 conference on
  Computational Models of Argument: Proceedings of COMMA 2010},
  \bibinfo{year}{2010}, pp. \bibinfo{pages}{3--12}.
\bibitem[{Prakken(2002)}]{prakken2002intuitions}
\bibinfo{author}{H.~Prakken},
\newblock \bibinfo{title}{Intuitions and the modelling of defeasible reasoning:
  some case studies},
\newblock \bibinfo{journal}{arXiv preprint cs/0207031}  (\bibinfo{year}{2002}).
\bibitem[{Horty(2001)}]{horty2001argument}
\bibinfo{author}{J.~F. Horty},
\newblock \bibinfo{title}{Argument construction and reinstatement in logics for
  defeasible reasoning},
\newblock \bibinfo{journal}{Artificial intelligence and Law}
  \bibinfo{volume}{9} (\bibinfo{year}{2001}) \bibinfo{pages}{1--28}.
\bibitem[{Caminada et~al.(2012)Caminada, Carnielli, and
  Dunne}]{caminada2012semi}
\bibinfo{author}{M.~W. Caminada}, \bibinfo{author}{W.~A. Carnielli},
  \bibinfo{author}{P.~E. Dunne},
\newblock \bibinfo{title}{Semi-stable semantics},
\newblock \bibinfo{journal}{Journal of Logic and Computation}
  \bibinfo{volume}{22} (\bibinfo{year}{2012}) \bibinfo{pages}{1207--1254}.
\bibitem[{Pollock(1995)}]{pollock1995cognitive}
\bibinfo{author}{J.~L. Pollock},
\newblock \bibinfo{title}{Cognitive carpentry, a blueprint for how to build a
  person},
\newblock \bibinfo{publisher}{Mit Press}, \bibinfo{year}{1995}.
\bibitem[{Rizzo et~al.(2018)Rizzo, Majnaric, Dondio, and
  Longo}]{rizzo2018investigation}
\bibinfo{author}{L.~Rizzo}, \bibinfo{author}{L.~Majnaric},
  \bibinfo{author}{P.~Dondio}, \bibinfo{author}{L.~Longo},
\newblock \bibinfo{title}{An investigation of argumentation theory for the
  prediction of survival in elderly using biomarkers},
\newblock in: \bibinfo{booktitle}{IFIP International Conference on Artificial
  Intelligence Applications and Innovations}, \bibinfo{organization}{Springer},
  \bibinfo{year}{2018}, pp. \bibinfo{pages}{385--397}.

\end{thebibliography}







\end{document}